\documentclass[12pt]{article}

\def\hybrid{\topmargin 0pt	\oddsidemargin 0pt 
	\headheight 0pt	\headsep 0pt
	\textheight 9in		% US paper
	\textwidth 6.1in	% A4 paper (old 6.25in)
	\marginparwidth .875in
	\parskip 5pt plus 1pt	\jot = 1.5ex}

\hybrid

\catcode`\@=11

\def\numberbysection{\@addtoreset{equation}{section}
	\def\theequation{\thesection.\arabic{equation}}}

\def\underline#1{\relax\ifmmode\@@underline#1\else
	$\@@underline{\hbox{#1}}$\relax\fi}

\def\titlepage{\@restonecolfalse\if@twocolumn\@restonecoltrue\onecolumn
	 \else \newpage \fi \thispagestyle{empty}\c@page\z@	
	\def\thefootnote{\fnsymbol{footnote}} }

\def\endtitlepage{\if@restonecol\twocolumn \else \newpage \fi
	\def\thefootnote{\arabic{footnote}} 
	\setcounter{footnote}{0}}  %\c@footnote\z@ }

\catcode`@=12
\relax

\numberbysection

\usepackage{graphicx}
\usepackage{latexsym}
\usepackage{amssymb}
\usepackage{amsopn}
\def\Mhalf{{\cal M}^{1/2}_g}
\def\Mhodd{{\cal M}^{1/2}_g}
\def\Mod{{\cal M}_{g}}
\def\Szbar{\bar{S}_0}
\def\Zsc{Z_{\rm sc}}

\def\azo{A^{0,1}}
\def\bbC{\mathbb{C}}
\def\bbR{\mathbb{R}}
\def\bbZ{\mathbb{Z}}

\def\cD{{\cal D}}

\def\cC{\mathcal{C}}
\def\cE{\mathcal{E}}
\def\cL{\mathcal{L}}
\def\cM{\mathcal{M}}

\def\cR{\mathcal{R}}
\def\cV{\mathcal{V}}
\def\chris#1#2{\stackrel{\circ}{\Gamma}\kern-0.5ex\vphantom{\Gamma}^{#1}_{#2}}
\def\dbar{\bar{\partial}}
\def\delX{\nabla_X^{0,1}}
\def\delzX{\stackrel{\circ}{\nabla}\vphantom{\nabla}_X^{0,1}}
\def\dirac{\dbar_{1/2}}
\def\dzbar{\partial_{\zbar}}
\def\dz{\partial_z}
\def\d{\partial}
\def\lamho{\Lambda^{1/2,1}}	%% lambda 1/2,1
\def\lamhz{\Lambda^{1/2,0}}	%% lambda 1/2,0
\def\lamzo{\Lambda^{0,1}}	%% lambda 0,1
\def\lamzz{\Lambda^{0,0}}	%% lambda 0,0

\def\mapsm{\Map(\Sigma,M)}
\newcommand{\mapsspinm}{\Map(\Sigma,\SPIN(M))}
\newcommand{\mapsspinn}{\Map(\Sigma,\Spin(n))}
\def\nablazX{\stackrel{\circ}{\nabla}\kern -0.5ex\vphantom{\nabla}_X}
\def\onep{\mathbf{1}^{\perp}}
\def\one{\mathbf{1}}
\def\psiz{\psi}	%% \def\psiz{\psi_\theta}
\def\scalar{\dbar_0}
\def\zbar{\bar{z}}
\newcommand{\Aoz}{A^{1,0}}
\newcommand{\Azob}{\overline{\Azo}}
\newcommand{\Azo}{A^{0,1}}
\newcommand{\BBox}{\raisebox{-0.25ex}{\large$\Box$}}

\newcommand{\Bzobp}{\overline{B'{}^{0,1}}}
\newcommand{\Bzob}{\overline{B^{0,1}}}
\newcommand{\Bzop}{B'{}^{0,1}}
\newcommand{\Bzo}{B^{0,1}}
\newcommand{\cMt}{\widetilde{\mathcal{M}}}
\newcommand{\dhalf}{\partial_{1/2}}
\newcommand{\Honehat}{\widehat{\Hone}}
\newcommand{\Hone}{H_1(\Sigma,\bbZ)}
\newcommand{\Hoz}{H^{1,0}(\Sigma)}
\newcommand{\Hzo}{H^{0,1}(\Sigma)}
\newcommand{\Lpz}{L_{P_0}}

\newcommand{\Loz}{\Lambda^{1,0}(\Sigma)}
\newcommand{\omegatil}{\tilde{\omega}}
\newcommand{\nutil}{\tilde{\nu}}
\newcommand{\Lzo}{\Lambda^{0,1}(\Sigma)}
\newcommand{\Lzz}{\Lambda^{0,0}(\Sigma)}
\newcommand{\Sigmat}{\widetilde{\Sigma}}
\newcommand{\Toz}{T^{1,0}(\Sigma)}
\newcommand{\binv}{\BBox_{-}^{-1}}
\newcommand{\cDt}{\widetilde{\mathcal{D}}}
\newcommand{\cLt}{\widetilde{\mathcal{L}}}
\newcommand{\db}{\bar{\partial}}
\newcommand{\detonep}{\det_{\onep}\BBox_-}
\newcommand{\detp}{\det_{\onep}\Delta_0}
\newcommand{\detprime}{\det_{\onep}}

\newcommand{\dyb}{\bar{\partial}_{Y}}
\newcommand{\dy}{\partial_{Y}}
\newcommand{\e}[1]{e^{\mbox{$\displaystyle#1$}}}
\newcommand{\half}{\frac{1}{2}}
\newcommand{\ie}{{\it i.e.\/}}

\newcommand{\zanalog}{\mathfrak{z}}
\newcommand{\ddelta}{\dbar_{\delta}}

\DeclareMathOperator{\Sym}{Sym}
\DeclareMathOperator{\norm}{norm}
\DeclareMathOperator{\ext}{ext}
\DeclareMathOperator*{\Tr}{Tr}
\DeclareMathOperator{\CP}{\mathbb{C}\mathbb{P}}
\DeclareMathOperator{\DET}{DET}

\DeclareMathOperator{\Index}{Index}
\DeclareMathOperator{\Map}{Map}
\DeclareMathOperator{\Met}{Met}
\DeclareMathOperator{\PF}{PF}

\DeclareMathOperator{\Sp}{Sp}
\DeclareMathOperator{\SO}{SO}
\DeclareMathOperator{\so}{\mathfrak{s}\mathfrak{o}}
\DeclareMathOperator{\SPIN}{SPIN}
\DeclareMathOperator{\Spin}{Spin}
\DeclareMathOperator{\Todd}{Todd}
\DeclareMathOperator{\cclass}{c}
\DeclareMathOperator{\ch}{ch}
\DeclareMathOperator{\coker}{coker}
\DeclareMathOperator{\ev}{ev}
\newcommand{\evtil}{\widetilde{\ev}}
\DeclareMathOperator{\hol}{hol}
\DeclareMathOperator{\mod}{mod}
\DeclareMathOperator{\pf}{pf}
\DeclareMathOperator{\pont}{p}

\DeclareMathOperator{\real}{re}
\DeclareMathOperator{\vol}{vol}
\def\bea{\begin{eqnarray}}
\def\beq{\begin{equation}}
\def\eea{\end{eqnarray}}
\def\eeq{\end{equation}}
\newcommand{\Tronep}{\Tr_{\onep}}
\newtheorem{theorem}[equation]{Theorem}
\newtheorem{corollary}[equation]{Corollary}
\newtheorem{lemma}[equation]{Lemma}
\newtheorem{proposition}[equation]{Proposition}
\providecommand{\href}[2]{#2}
\begin{document}
\pagenumbering{roman}

\begin{titlepage}
\strut\par\noindent
April 2001\hfill UMTG--228\newline
\strut\hfill \texttt{hep-th/0104199}

\vspace{.5in}
\begin{center}
{\bf\Large Beyond the Elliptic Genus}\\[.75in]
{\bf Orlando Alvarez\footnote{email: 
\href{mailto:alvarez@physics.miami.edu}{\texttt{alvarez@physics.miami.edu}}
}}\\
{\it Department of Physics}\\
{\it University of Miami}\\
{\it P.O. Box 248046}\\
{\it Coral Gables, FL 33146 USA}\\
and\\
{\bf I. M. Singer\footnote{email: 
\href{mailto:ims@math.mit.edu}{\texttt{ims@math.mit.edu}}
}}\\
{\it Department of Mathematics}\\
{\it Massachusetts Institute of Technology}\\
{\it 77 Massachusetts Avenue, Rm. 2--387}\\
{\it Cambridge, MA 02139 USA}
\end{center}
\vspace{.5in}
\begin{abstract}
Given a Riemann surface $\Sigma$  and a riemannian
manifold $M$ with certain restrictions, we construct a cobordism
invariant of $M$. This invariant is a generalization of the
elliptic genus and it shares some similar properties. 
\end{abstract}
\end{titlepage}

\small
\tableofcontents
\normalsize
\newpage
\pagenumbering{arabic}

\section{Introduction}

The analytic index of an elliptic differential operator $D:
C^\infty(E_+) \to C^\infty(E_-)$ between vector bundles over a 
manifold $M$ is  defined as
\beq
	\Index(D) = \dim\ker D - \dim\coker D^{*}\;.
\eeq
This index can be computed  by using heat 
evolution operators. There are two natural laplacians associated 
with the elliptic operator $D$:
\bea
	\Delta_+ &=& D^* D : C^\infty(E_+) \to C^\infty(E_+)\;, \\
\Delta_- &=& D D^* : C^\infty(E_-) \to C^\infty(E_-)\;.  \eea The
index may be expressed as \beq \Index(D) = \Tr \exp(-t\Delta_+) - \Tr
\exp(-t \Delta_-)\;, t>0, \eeq because if $\Delta_+ \phi =
\lambda\phi$, then $\Delta_-(D\phi)= \lambda (D\phi)$.  Thus all but
the $1$~eigenvalues of $e^{-t\Delta_+}$ cancel those of
$e^{-t\Delta_-}$ in the difference of the traces.  The cancellation is
incomplete because one cannot identify $\ker D$ with $\ker D^*$. 
Using heat operators leads one to attempt to find a classical quantum
mechanics problem with the laplacians being the hamiltonians.  The end
result is supersymmetric quantum mechanics which has a supersymmetric
path integral formulation
\cite{Alvarez-Gaume:1983at,Friedan:1984xr,Witten:AS}.

We can go beyond quantum mechanics to quantum field theory and ask
whether there are generalizations of the index theorem.  In the
context of $(1+1)$ dimensional quantum field theories, the answer is
yes in the form of the elliptic genus
\cite{Alvarez:1987de,Alvarez:1987wg,Pilch:1987en,Witten:1987bf,Witten:elliptic}. 
For a survey of the mathematical literature look in \cite{Landweber:elliptic}.
The $(1+1)$ dimensional field theory is formulated on a torus where
there is a notion of a hamiltonian, \ie, laplacian.  Formally, the
elliptic genus is the $S^1$-index of a formal differential operator,
the Dirac-Ramond operator, the Dirac operator with potential function
Clifford multiplication by $x'(\sigma)$ on the loop space $LM$.  If
one goes beyond genus one the Hamiltonian interpretation is lost, but
there is still a path integral.  Can one make any sense of this path
integral as some generalization of the index and what is it?  In this
paper we show that the semiclassical approximation of the path
integral gives a cobordism invariant generalizing the elliptic 
genus to the case of genus $g>1$.

In Section~\ref{sigma-model} we review the supersymmetric sigma model,
its action and partition function. We explain in
Section~\ref{susy-path-integral} conditions needed on the target
manifold to cancel anomalies and make the path integral formally well
defined. We derive the semiclassical approximation in
Section~\ref{sc-section}, obtaining the semiclassical partition
function as a ratio of determinants in (\ref{Z-sc}). The semiclassical
limit is ``topological'' but not a topological quantum field theory
\cite{Witten:1988xj,Atiyah:topo}.

We derive a differential equation for one of these determinants in
Section~\ref{sec-diff} and use it to ultimately compute the
determinant in terms of $\vartheta$-functions whose characteristics
are determined in Appendix~\ref{app-kappa}.  We identify the
$\vartheta$-function with a cross section of our determinant line
bundle in Section~\ref{hol-triv}.  Using an explicit construction of
the determinant line bundle (Section~\ref{detd-construction}) we
compute $\detonep$ in Section~\ref{detM-section}
(Theorem~\ref{thm-detonep}). In Section~\ref{final-section} we 
discuss our final formula which gives a cobordism invariant 
generalizing the elliptic genus as we explain. There we obtain a simple 
``relative invariant'' by taking ratios.

In the appendices we tried to make explicit known material in 
algebraic geometry.

The body of the paper was written over two years ago.  We had hoped to
exhibit our semiclassical partition function explicitly as a
nonholomorphic section of a holomorphic line bundle over spin moduli
space.  We did not succeed in doing so.  In the meantime line bundles
over jacobians have received considerable attention because of
M-theory (partition functions for self-dual fields and chiral
anomalies).  Though we are aware of some of these developments
\cite{Freed:2000tt,Hopkins:2000unpub}, we have not incorporated their
viewpoint into our computations.

\section*{Acknowledgments}

We would like to thank D. Freed, Joe Harris, M.J. Hopkins, T. Ramadas, and Paul
Windey for illuminating discussions.  The work of O.A. was supported
in part by National Science Foundation grant PHY--9870101.

\section{The Supersymmetric Sigma Model}
\label{sigma-model}

\subsection{The bosonic model} 

Let $\Sigma$ be a Riemann surface of genus\footnote{We use $g$ for 
both the genus of $\Sigma$ and the metric on $M$.} $g$ and let $M$ be a 
connected and oriented riemannian manifold of dimension $d$.  Consider 
a map $X:\Sigma\to M$; then the classical nonlinear sigma model is 
defined by the ``energy'' action 
\beq
\label{nlsm-action} 
I(X) = \half \left\langle dX, dX \right\rangle\;, 
\eeq 
where $dX$ denotes the 
differential map $dX: T\Sigma \to TM$.  The natural inner product 
induced by the riemannian structures is denoted by angular brackets.  
If $(x^1,x^2)$ are real
local coordinates\footnote{Our notation is 
that $d^2 x = dx^1 \wedge dx^2$.} on $\Sigma$ and if we abuse notation 
and also denote local coordinates on $M$ by $X$, 
action~(\ref{nlsm-action}) may be written in the form 
\bea
I(X) &=& 
\half \int_\Sigma d^2 x\; \sqrt{\det \gamma}\; \gamma^{ab}(x)\, 
g_{\mu\nu}(X(x)) {\d X^\mu \over \d x^a}{\d X^\nu \over \d x^b}\\
&=& i \int_\Sigma g_{\mu\nu}(X(x)) \d X(x)^\mu \wedge \dbar X(x)^\nu\;.
\eea
In the first line, $\gamma_{ab}(x) dx^a\otimes dx^b$ is
the riemannian metric on $\Sigma$, and $g_{\mu\nu}(X)
dX^\mu\otimes dX^\nu$ is the riemannian metric on $M$. In the 
second line we have exploited the complex structure on $\Sigma$ 
induced by its riemannian structure to write the action in a way 
that  depends manifestly only on the complex structure.

\subsection{The supersymmetric model}

The chiral Dirac operator on a Riemann surface $\dirac: \lamhz(\Sigma) 
\to \lamho(\Sigma)$ has numerical index zero.  It 
is also a skew symmetric operator and consequently has a $\mod 2$ index, 
$(\dim \ker \dirac) \mod 2$ which is a topological invariant.  An even 
spin structure is one where $\dim \ker \dirac = 0\mod 2$ and an odd 
spin structure is one where $\dim \ker \dirac = 1\mod 2$, see 
\cite{Atiyah:spin}.

The supersymmetric version of action~(\ref{nlsm-action}) requires the
introduction of a fermionic field $\psi$ which is a section of the
bundle $K^{1/2} \otimes X^*(TM)$ where $K$ is the canonical bundle on
$\Sigma$. We have to pick a square root of the canonical bundle, \ie,
a spin structure $s$ on $\Sigma$. Later we will see that we 
have to pick an odd spin structure. Let $\nablazX$ be the induced
Riemannian
covariant differential\footnote{The $X$ subscript is introduced to
emphasize that the operator depends on the map $X$.} on the bundle
$K^{1/2} \otimes X^*(TM)$. If we use the complex structure on $\Sigma$
and decompose the tangent bundle as $T\Sigma = T^{1,0}\Sigma \oplus
T^{0,1}\Sigma$ then the differential has a natural decomposition as
$\nablazX = \nablazX^{1,0} + \nablazX^{0,1}$. In local coordinates, 
the covariant derivative of the section
\beq
	\psiz^\mu \; \sqrt{dz} \otimes {\d\over \d X^\mu}\;,
\eeq
is given by
\beq \label{cov-Riem}
\left(\nablazX^{0,1}\psi\right)^\mu = \dbar\psi^\mu 
 +(\dbar X^\lambda)\,\chris{\mu}{\lambda\nu}(X) \psi^\nu\;,
\eeq
where $\chris{\mu}{\lambda\nu}$ are the Christoffel symbols for the 
Riemannian connection on $TM$. 

The supersymmetry transformation laws are
\bea
	\delta X &=& \epsilon \psiz \\
	\delta \psiz &=& \epsilon\dz X \;,
\eea
where $\epsilon$ is a local holomorphic section\footnote{There are no 
global holomorphic sections for genus $g>1$.  This problem 
will be addressed shortly.} of 
$K^{-1/2}$.  For the mathematicians, $\delta X$ should be interpreted 
as a tangent vector to the space of maps $\mapsm$, i.e., a cross 
section of $X^*(TM)$.
The {\it naive\/} supersymmetric action may be written in 
local complex coordinates as
\beq  \label{susy-action}
	I = I_g + I_B \;,
\eeq
where
\bea
I_g &=& i \int_\Sigma dz\wedge d\zbar\; g_{\mu\nu}(X(z))
\dz X^\mu \dzbar X^\nu \nonumber\\
&+& i \int_\Sigma dz\wedge d\zbar\; g_{\mu\nu}(X)
\left( \dzbar\psiz^\mu + (\dzbar 
X^\lambda)\chris{\mu}{\lambda\rho}(X) \psiz^\rho \right) 
\psiz^\nu\;,\\
&=& i \int_\Sigma g_{\mu\nu}(X)\left( \d X^\mu \wedge \dbar X^\nu +
	(\delzX\psi)^\mu \psi^\nu \right)\; ;
\eea
and
\beq
	I_B = i\int_\Sigma dz \wedge d\zbar
		\left(B_{\mu\nu}(X) \dz X^\mu \dzbar X^\nu
		+ \half C_{\lambda\mu\nu}(X)\psiz^\lambda \psiz^\mu \dzbar X^\nu 
		\right)\;.
		\label{Ib-def}
\eeq For the moment, we take $B=\half B_{\mu\nu}dX^\mu\wedge dX^\nu$
to be a real $2$-form\footnote{$B$ is not really a $2$-form, see the
discussion in \cite{Alvarez:1985es,Freed:2000tt,Hopkins:2000unpub}.}
on $M$ with \beq C =dB= \frac{1}{3!}\; C_{\mu\nu\rho}\;dX^\mu \wedge
dX^\nu \wedge dX^\rho\;.  \eeq We will be more precise later on the
exact interpretation of the $B$ term.  For now suffices to say that it
is required for anomaly cancellation.

The action is not invariant under supersymmetry in genus larger 
than one. For example,
the  transformation law for action $I_g$ under the 
supersymmetry transformation is 
\beq
	\delta I_g = \int_\Sigma 
	\epsilon \dz\left( g_{\mu\nu}(X)\psiz^\mu \dzbar 
	X^\nu \right)\;.
\eeq For the supersymmetry to be a symmetry of this action, $\epsilon$
must be holomorphic and this only happens in genus zero or genus one.
Note that in genus one, a constant $\epsilon$ tells us that $\psi$
must belong to an odd spin structure,  a consequence of the
SUSY transformation laws and the periodicity of $X$ around any cycle
in $\Sigma$. Therefore the field $\psiz$ is periodic on a torus,
the boundary condition that is consistent with supersymmetry. For
genus $g>1$, there is no global supersymmetry and one can only talk
about supersymmetry locally. The  classical holomorphic
supercurrent is  of type
$(3/2,0)$.

The above action defines a sensible classical conformal
field theory.  Quantum mechanically, this is not so.  Firstly, there
are global fermionic anomalies as discussed in
\cite{Moore:1984dc,Moore:1985ws} and local Adler-Bell-Jackiw anomalies
associated with gauge transformations in $TM$.  The $I_B$ term is used
to eliminate these anomalies.  Secondly, due to the conformal anomaly,
the above is not in general a conformal field theory.  But we will
only use the semiclassical approximation about a constant background,
which is a conformal field theory.

\section{The Supersymmetric Path Integral}
\label{susy-path-integral}

\subsection{Full theory}

The path integral for action~(\ref{susy-action}) involves 
integrating over all maps from $\Sigma$ to $M$ and integrating over 
all fermionic sections of $K^{1/2}\otimes X^*(TM)$. Since the 
fermions enter quadratically, we can perform the fermionic integral 
obtaining the following formal expression for the partition section
\bea 
Z(g_{\mu\nu},\gamma_{ab},s) &=& \int\limits_{\mapsm} [\cD X]\;\pf(\delX)\;
	\nonumber\\
	\label{Z-exact}
	&\times&\exp\left\{
	i\int_\Sigma dz \wedge d\zbar\;
	\left[ g_{\mu\nu}(X(z))+B_{\mu\nu}(X(z)) \right] 
	\dz X^\mu \dzbar X^\nu \right\}\;.
\eea
In the above, $\delX$ is defined just like (\ref{cov-Riem}) except 
that the connection coefficients are given by
\beq \label{Gamma-full}
	\Gamma^\lambda_{\mu\nu} = \chris{\lambda}{\mu\nu} + \half 
	C_\mu{}^\lambda{}_\nu\;.
\eeq
This is a metric compatible connection with torsion.
In the expression for the partition section~(\ref{Z-exact}),
$\pf(\nabla_X^{0,1})$ is the pfaffian section of the
pfaffian line bundle $\PF(\nabla_X^{0,1})$. We  emphasize that
the partition section depends on the metric on the target space $M$,
the metric on the Riemann surface $\Sigma$, and the spin structure 
$s$ of the Riemann surface.

The expression for the partition section $Z$ may be interpreted as an
averaging of the pfaffian section over $\mapsm$.  This can be done
only if the pfaffian line bundle is a trivial line bundle over $\mapsm
$.  If not we have an anomaly in the sigma model as discussed in
\cite{Moore:1984dc,Moore:1985ws}.  General arguments tell us that the
determinant line bundle $\DET(\delX) \to \mapsm$ exists over compact
sets of $\mapsm$.  On such sets it has a canonical section
$\det(\delX)$ because $\Index \delX=0$.  Freed's Theorem~3.1
\cite{Freed:1986hv}
guarantees that there exists a line bundle $\PF(\delX)$ with a
canonical isomorphism $\PF(\delX)\otimes\PF(\delX) = \DET(\delX)$ and
with a canonical section $\pf\delX$ such that $(\pf\delX)^{\otimes 2}
= \det(\delX)$.  Since the pfaffian line bundle $\PF(\delX) \to
\mapsm$ must be trivial to prevent the anomaly, the line bundle
$\DET(\delX)\to \mapsm$ must also be trivial.  The family's index
theorem  shows that the first Chern
class of the determinant line bundle is given by 
\begin{equation}
\cclass_1(\DET(\delX)) = -\half \int_\Sigma \ev^* [\pont_1(M)]\;, 
\end{equation}
where $\ev : \Sigma \times \mapsm \to M$ is the evaluation map.  The
cohomology class $2\pi\cclass_1(\DET(\delX))$ may be represented by a
curvature $2$-form $\mathcal{F}$ of the determinant line bundle
$\DET(\nabla_X^{0,1})$ (which comes equipped with a Quillen
connection).  $\mathcal{F}/2\pi = -\half \int_{\Sigma}
\ev^{*}\pont_{1}(M)= -\half \int_{X(\Sigma)}\pont_{1}(M)$ at
$X\in\mapsm$, see Bismut and Freed \cite{Bismut:58110a,Bismut:58110b}. 
We assume that $\dim M$ is even and greater than $2$.  We also assume
that $\pi_{j}(M)=0$ for $j \le 3$.  The condition $\pi_{3}(M)=0$
greatly simplifies the analysis.  We discuss complications when
$\pi_{3}(M)\neq 0$ later.  Our assumptions on $M$ imply that $M$ is an
oriented, connected spin manifold and that $\pi_{j}(\mapsm) =0$ for
$j=0,1$.  Hence $\PF(\nabla_X^{0,1})$ is the \emph{unique} square root
of $\DET(\nabla_X^{0,1})$ over $\mapsm$.  We now assume that
$[\pont_1(M)]=0$, then $\DET(\delX) \simeq \mapsm\times\bbC$, i.e.,
$\DET(\delX)$ is isomorphic to a trivial $C^\infty$ line bundle.

Triviality of the line bundle is not sufficient.  A
locality requirement in physics necessitates that there be no local
anomaly.  Counterterms have to be added to cancel the local
obstructions and not just the topological ones; see
Section~\ref{sec-anomaly}.

\subsection{Determinant and pfaffian line bundles}

The twisted chiral Dirac operator $\delX : \lamhz(\Sigma,X^*(TM)) \to 
\lamho(\Sigma,X^*(TM))$  has numerical index zero.  It has a 
determinant line bundle $\DET(\delX)$ with canonical section 
$\det\delX$ over the parameter space $\mapsm 
\times\Met(M)\times\Mhalf$.  Here $\Met(M)$ is the space of metrics on 
$M$ and $\Mhalf$ is odd spin moduli space for genus $g$ Riemann surfaces.  
Actually, the operator $\delX$ depends on a choice of orthogonal 
connection $A$ in $\mathcal{A}$, the space of orthogonal connections;
thus the parameter space is really $\mapsm\times\mathcal{A} 
\times\Met(M)\times\Mhalf$.

Note that the vector bundle $X^*(TM)$ is a real vector bundle and so 
$\delX$ has a $\mod 2$ index.  The $\mod 2$ index theorem states among 
other things that
\begin{theorem} 
	\label{mod2-index}
	$$\dim\ker\delX = (\dim M)(\dim\ker\dirac) \mod 2\;.$$
\end{theorem}
(because $X^*(TM)$ is stably trivial as a real bundle over 
$\Sigma$). Hence, if $M$ is even dimensional, $\dim\ker\delX = 0 \mod 2$.

We restrict ourselves to odd spin structures; the basic Dirac operator 
has a zero mode so that ordinarily the Pfaffian and the path integral 
would vanish.  However we have twisted by the pullback bundle 
$X^*(TM)$ and $\dim M$ is even so the $\dim\ker\delX=0\mod 2$.  
Generically, we do not have a zero mode and this fact allows for a 
nonvanishing partition function.  Even spin structures do not give 
topological invariants of $M$ in the semiclassical approximation.

\subsection{Anomaly cancellation}
\label{sec-anomaly}

In this section we address two issues.  First we give an explicit
(local) trivialization of the line bundle $\PF(\delX)$ so that the
section $\pf(\nabla_X^{0,1})$ becomes a function.  Second, we show how
$I_{B}$ cancels the gauge anomaly of $I_{g}$ (general considerations
\cite{Bardeen:1984pm} imply that consequently there will be no
gravitational anomaly as well).

We first give a rough outline of the chain of arguments which leads to 
anomaly cancellation.  For the moment we do not worry about 
normalizations factors since later we will do it more carefully.

We have a map $X:\Sigma\to M$.  The Dirac operator $\delX$ involves 
the connection on the pullback bundle $X^*(TM)$.  If $v$ is an 
infinitesimal gauge transformation then the anomaly is given (up to 
factors of $\pi$ and integers) by
$$
	\left(\pf(\delX)\right)^{-1}\delta \left(\pf(\delX)\right) =
	\int_\Sigma \Tr(vdA)\;,
$$
where we have denoted the pullback connection by $A$. 

Assume we have a two form $B$ on 
$M$. Action~(\ref{Ib-def}) contains a term
$$
	I' = \int_\Sigma \left(X^* B\right)\;.
$$
If we can arrange that under a gauge transformation the 
variation in $I'$ is given by
$\delta I' = - \int_\Sigma \Tr(vdA)$ then we can cancel the anomaly 
in the pfaffian with the $B$ term.

The local cancellation of the anomaly is based on the observation that 
if $A$ is a connection, then under an infinitesimal gauge 
transformation $A \to A + dv + [A,v]$ the Chern-Simons form $\alpha = 
\Tr(A dA + \frac{2}{3}A^3)$ transforms as $\alpha \to \alpha + 
d\Tr(vdA)$.  The object is to use the Chern-Simons form to cancel the 
anomaly.  Since $\pi_1(M)=\pi_2(M)=0$, the image 
$X(\Sigma)$ must be a boundary.  There exists $N_X \subset M$ such 
that $\partial N_X = X(\Sigma)$.  Thus we conclude that
$$
	\delta I' = - \int_\Sigma \Tr(vdA) = -\int_{N_X} d\left(\Tr vdA\right)
	=-\int_{N_X} \delta\alpha=-\delta\int_{N_X} \alpha\;.
$$
We require
$$
	I' = \int_\Sigma X^*(B) = -\int_{N_X} \alpha +\mbox{constant}\;.
$$
Since $\pont_1(M)=0$ we have $d\alpha=0$ and thus $\alpha$ has a local 
antiderivative and we want this
antiderivative to be $B$ justifying the equation above.

We now give the argument more carefully.  Since $[\pont_{1}(M)]=0$, we
can choose a $3$-form $H$ on $M$ such that\footnote{Just as the
$4$-form $\pont_{1}(M)$ depends on an $\SO(n)$ connection $A$ for a
given metric, so does $H$.} $dH=\pont_{1}(M)$.  So $\mathcal{F} =
d\omega$ where $\omega$ is the $1$-form on $\mapsm$ equal to
$-2\pi\times\half\int_{X(\Sigma)}H$ at $X\in\mapsm$.  Hence
$\DET(\nabla_X^{0,1})$ is trivialized by using the connection
$0+\omega$; the line bundle $\PF(\nabla_X^{0,1})$ is then the trivial
bundle with connection $0 + \half \omega$, with $\half \omega$ at
$X\in\mapsm$ equal to $-2\pi\times\frac{1}{4}\int_{X(\Sigma)}H$ [Note that the
cohomology class $\pont_{1}(M)/4$ is integral because $M$ is spin]. 
Thus $\pf(\nabla_X^{0,1})$ is now a function on $\mapsm$.  Another
choice $H'=H + db$, $b$ a $2$-form, would give the connection
$0+\omega'$ with $\omega'=\omega + d\mu$ where $\mu$ is the function
on $\mapsm$ given by $X\mapsto \int_{X(\Sigma)}b$.

To study the gauge anomaly, we let $\alpha=\alpha(A)$ be the
Chern-Simons form on $\SPIN(M)$
for the connection $1$-form $A$, so that $d\alpha = \frac{1}{8\pi^{2}}
\Tr\Omega^{2} = \pi^{*}(\pont_{1}(M))$.  Here $\Omega$ is the
$\so(n)$ valued curvature $2$-form on $\SPIN(M)$.  Let $C= \alpha -
\pi^{*}H$; hence $dC=0$.  The homotopy exact sequence for the
principal bundle $\pi:\SPIN(M)\to M$ and $\pi_{j}(M)=0$ for $j \le 3$
imply that $\pi_{3}(\SPIN(M)) \cong \pi_{3}(\Spin(n)) = \bbZ$. 
Since $\pi_{j}(\SPIN(M))=0$ for $j=1,2$, we get that
$H_{3}(\SPIN(M),\bbZ)\cong\pi_{3}(\SPIN(M)) =\bbZ  $.  But the integral of
$C$ over a fundamental $3$-cycle in a fiber is $1$;  hence
$C$ represents a generator of
$H^{3}(\SPIN(M),\bbZ)$.

Note that $\mapsspinm$ is a principal bundle over $\mapsm$ with group 
$\mapsspinn$. Let $\evtil: \Sigma \times \mapsspinm \to \SPIN(M)$ be the 
evaluation map. Then $\evtil^{*}(C)$ is a closed integral $3$-from 
on $\Sigma\times \mapsspinm$ and $\int_{\Sigma}\evtil^{*}(C)$ is a 
closed $1$-form $\omegatil$ on $\mapsspinm$. We define a function 
$\cE(C)$ on $\mapsspinm$ as follows. Fix a trivial map 
$\widetilde{X}_{0}:\Sigma \mapsto P_{0}\in \SPIN(M)$. Let $\gamma$ be 
any path from $\widetilde{X}_{0}$ to $\widetilde{X}\in\mapsspinm$ so 
that $\gamma: [0,1]\times\Sigma \to \SPIN(M)$ with $\gamma(1) = 
\widetilde{X}$. Such a path exists because $\pi_{2}(\SPIN(M))=0$. 
Define
$$
    \cE(C)(\widetilde{X}) = \exp\left( 2\pi i \int_{\gamma([0,1] \times 
    \Sigma)} 
    C\right)
	= \exp\left(2 \pi i \int_{\gamma([0,1])} \tilde{\omega} \right)\,.
$$
Now $\cE(C)(\widetilde{X})$ is independent of the path $\gamma$; for 
if $\gamma_{1}$ is another such path then 
$\gamma_{1}^{-1}\gamma$ is a map of $S^{1}\times\Sigma \to 
P_{0}$ and
$$
    \int_{(\gamma_{1}^{-1}\gamma)(S^{1}\times\Sigma)} C =
    \int_{\gamma(S^{1}\times\Sigma)} C -
    \int_{\gamma_{1}(S^{1}\times\Sigma)} C
$$	
is an integer, i.e., $\tilde{\omega}$ represents an integral 
$1$-cocycle. 

Put another way, $0 + \tilde{\omega}$ is a connection on the trivial 
line bundle over $\mapsspinm$ with $0$ curvature and trivial holonomy 
for all closed paths. 
So $\tilde{\omega}$ is a pure gauge generating the gauge 
transformation $\cE(C)$.

We can define the function $\cE(C)$ a bit more abstractly.  Since $C$
represents an element of $H^{3}(\SPIN(M),\bbZ)$, there exists a
Cheeger-Simons differential $2$-character $B$ with $dB=C$.  Then
$\evtil^{*}B$ is a differential $0$-character on $\mapsspinm$ with
values in $S^{1}$ whose ``differential'' is $\tilde{\omega}$. 
$\evtil^{*}B$ is our function $\cE(C)$.  Changing $H$ to $H + db$,
changes $B$ to $B+b$.  The bosonic part of the action $I_{B}$ is to be
interpreted as $\log\cE(C)$.

Let $\cC : \mapsspinm \times \mapsspinn \to S^{1}$ be the function
$\cC(\widetilde{X},\phi) = \cE(\phi\cdot\widetilde{X})
\cE(\widetilde{X})^{-1}$.  It is a cocycle, i.e.,
$\cC(\widetilde{X},\phi)\cC(\phi\cdot \widetilde{X},\psi) =
\cC(\widetilde{X},\psi\phi)$.  The cocycle $\cC$ defines a circle
bundle $\mapsspinm \times_{\cC} S^{1}$ over $\mapsspinm$, the
equivalence relation is $(\widetilde{X}, z) \approx (\phi
\cdot\widetilde{X}, \cC(\widetilde{X},\phi)z)$.

This circle bundle comes equipped with a connection as follows. The 
trivial circle bundle $\mapsspinm \times S^{1}$ has connection 
$1$-form $d\theta - \tilde{\omega}$, i.e., the trivial connection 
$d\theta$ modified by the $1$-form $-\tilde{\omega}$. The connection 
descends to $\mapsspinm \times_{\cC} S^{1}$ if $d\theta - 
\tilde{\omega}$ is invariant under the map $(\widetilde{X},z) 
\mapsto (\phi\cdot\widetilde{X}, \cC(\widetilde{X},\phi)z)$. 
Equivalently, we must show that 
$\mathcal{L}_{V} (d\theta-\tilde{\omega}) =0$ where $\mathcal{L}_{V}$ 
is the Lie derivative with respect to the vector field generating the 
$1$-parameter family of maps given by $\phi_{t}= e^{tf}$, 
$f:\Sigma\to\so(n)$. Since $d\theta -\tilde{\omega}$ is closed, we 
need
only show that $d\theta(V) -\tilde{\omega}(V)$ is constant. But $V$ in 
the $\mapsspinm$ component at $\widetilde{X}$ is the vector field 
along $\widetilde{X}(\Sigma)$ equal to $f(\Sigma)$, i.e., $V$ at 
$\widetilde{X}(\sigma)$ is the vertical vector $f(\sigma)$. 
Consequently $\tilde{\omega}(V)$ at $\widetilde{X}$ is
$$
    \int_{\widetilde{X}}C \left(\frac{\partial}{\partial z}, 
	\frac{\partial}{\partial \zbar}, f(\sigma) \right)\,.
$$
The component of $V$ in the $S^{1}$-direction can be computed in the 
following fashion. We have a $1$-parameter family of maps $e^{2\pi i 
\theta} \to e^{2\pi i(\theta + g(t))}$ where
$$
    g(t) = \int_{\phi_{t}(\ext \widetilde{X})} C
	 - \int_{\ext \widetilde{X}} C
$$
and $\ext \widetilde{X}$ is an extension of $\widetilde{X}$ to a map 
of a three manifold $N$ with boundary $\Sigma$. The map 
$\ext\widetilde{X}: N \to \SPIN(M))$ with $\ext\widetilde{X}|_{\Sigma} = 
\widetilde{X}$ induces a map $\ext\phi_{t}\widetilde{X} : N_{t} \to 
\SPIN(M)$ with $\ext\phi_{t}\widetilde{X}|_{\partial 
\phi_{t}\widetilde{X}} = \phi_{t}\widetilde{X}$. 
$\phi_{t}(\ext\widetilde{X})$ pushes $\ext\widetilde{X}$ in the vertical 
direction determined by $f(t)$.

Thus the component of $V$ in the $\partial/\partial\theta$ direction
is $dg/dt|_{t=0}$ which equals
$\int_{\widetilde{X}}C(\frac{\partial}{\partial z},
\frac{\partial}{\partial \zbar}, f(\sigma))$.  Thus the circle
bundle
$\mapsspinm \times_{\cC} S^{1}$ over $\mapsspinm$ has the connection
$d\theta-\tilde{\omega}$ pushed down to it.  The curvature of this
connection is zero; in fact $d\theta-\tilde{\omega}$ is a closed form
representing an integral cohomology class.  So all holonomies are
trivial and this circle bundle can be trivialized using the
connection $d\theta-\tilde{\omega}$.  Let $\mathcal{T}$ be the
associated trivial line bundle.
Note that the line bundle $\mathcal{T}$ has a natural nonvanishing section $s$ 
that is the descendant of the function
$\tilde{s}(\widetilde{X},z) = \cE(C)(\widetilde{X})z$ on $\mapsspinm 
\times \bbC$. So $s^{1/2}$ is a nonvanishing section of the line 
bundle $\mathcal{T}^{1/2}$; since $\mathcal{T}^{1/2}$ has been 
trivialized, $s^{1/2}$ is a function.

Implicit in our construction of the line bundle $\mathcal{T}$ is its dependence 
on the spin connection $A\in \mathfrak{A}$, the set of all 
connections on $\SPIN(M)$. So just as $\DET(\delX)$ is a line bundle 
over $\mapsm \times \mathfrak{A}$, so is $\mathcal{T}$.

\begin{theorem}
	$\pf(\delX) \times s^{1/2}$ over $\mapsm \times \mathfrak{A}$ is
	invariant under the group of gauge transformations $\mathfrak{G}$
	on $\mathfrak{A}$.
\end{theorem}
The proof uses two lemmas.  Let $\phi$ be a gauge transformation on
$\SPIN(M)$ and let $\phi\cdot A$ denote its action on $A$.
From \cite{Freed:1995vw} we see that
$$
\cE(C)(\widetilde{X},\phi\cdot A)= \cE(C)(\widetilde{X},A)e^{-2\pi i
\int_{\ext\widetilde{X}} \Tr(\phi^{-1}d\phi)^3} e^{-2\pi i
\int_{\Sigma}  \widetilde{X}^{*}\Tr(A \wedge\phi^{-1}d\phi)}
$$
and thus we conclude that
\begin{lemma}
    $\tilde{s}(\widetilde{X}, \phi\cdot A, z) = \cE(C)(\widetilde{X}, 
    \phi\cdot A)z$ equals $e^{-2\pi i \{\quad\}}$ where
	$$ 
	\{\quad\} = \int_{\ext\widetilde{X}}\Tr(\phi^{-1}d\phi)^{3}
	+ \int_{\Sigma} \widetilde{X}^{*}\Tr(A \wedge \phi^{-1}d\phi)\,.
	$$
\end{lemma}
It is well known \cite{Bardeen:1969md} that the change of the fermions
determinant under a gauge transformation is given by the non-abelian
anomaly\footnote{The path integral viewpoint is due to Fujikawa
\cite{Fujikawa:1979ay}.  For a review using more modern geometrical
language look at \cite{Alvarez-Gaume:1985dr}.}:
$$
\pf(\delX)(\phi\cdot X^*(A)) = \pf(\delX)(X^*(A)) e^{2\pi i
\{\quad\}}
$$
with 
$$
\{\quad\} = \int_{\ext\widetilde{X}}\Tr(\phi^{-1}
d\phi)^3 + \int_{\Sigma} X^*(\Tr A\wedge
\phi^{-1} d\phi).  
$$
We have that
\begin{lemma}
	$\pf(\delX)(X^{*}\phi\cdot A) = e^{2\pi i \{\!\{\quad\}\!\}}$ where
	$\{\!\{\quad\}\!\} = \half \{\quad\}$.
\end{lemma}

Concluding Remark: When $\pi_{3}(M) \neq 0$, $\DET(\delX)$ does not 
have a unique square root because $\mapsm$ is not simply connected. To 
see which square root $\pf(\delX)$ is, requires the K-theory formula 
(as opposed to a cohomology formula) for the pfaffian line bundle 
which in general is nonlocal \cite{A-S:5}. We do not address this 
problem here.

\subsection{Interpretation}

We emphasize that for genus $g>1$, the partition section does not have
an interpretation as the index of an operator.  We have neither an
$S^1$-index interpretation nor a topological quantum field theory
(TQFT) interpretation as in \cite{Witten:1988xj,Atiyah:topo}.  However,
we make the following observation.  Rescale the metric $g_{\mu\nu}(X)
\to g_{\mu\nu}(X)/\hbar$ and study the behavior of $Z$ as $\hbar\to
0$.  In this case the partition function has an asymptotic expansion
of the form 
\beq 
Z(g_{\mu\nu}/\hbar,\gamma_{ab},s) \sim \Zsc(g_{\mu\nu},\gamma_{ab},s)
+ O \left(\hbar^{1/2}\right)\;,
\eeq 
where the semi-classical partition
section $\Zsc$ is independent of $\hbar$.

Though we can say very little about the properties of the exact
partition section, we show that $\Zsc(g_{\mu\nu},\gamma_{ab},s)$ makes
sense and is a cobordism invariant of $M$ in the form of a
non-holomorphic section of a certain holomorphic line
bundle
over odd spin Teichmuller space.  Thus the
$(1,0)$ ``supersymmetric'' sigma model is a quantum field theory with
the property that its semi-classical limit gives topological
invariants but it is not a TQFT in the traditional sense.

\section{Semiclassical approximation} 
\label{sc-section}

The partition section may be written as \beq Z(g/\hbar,\gamma,s) = 
\int [\cD X][\cD \psi] \exp(-I(g,\gamma,s)/\hbar)\;, \eeq where 
$I(g,\gamma,s)$ is given by equation~(\ref{susy-action}).  As 
$\hbar\to 0$, we use the steepest descent approximation to evaluate 
the above.  This entails finding the critical points of the action.  
An example of a critical point is a constant map $X(z) = X_0$, and a 
holomorphic $\psiz$.  These are the global minima as far as the 
bosonic degrees of freedom.  From now on we neglect all other critical 
points.  Note that for a constant map, the pullback bundle $X^*_0(TM) 
= \Sigma\times TM_{X_0} = \Sigma\times\bbR^{2n}$ and therefore it 
makes sense for the section $\psiz$ to be holomorphic.  We assume we 
are working at a generic point in $\Mhodd$ where $\dim\ker\dirac=1$.

Next, we look at the quadratic fluctuations of the action. It is 
convenient to choose a Riemann
normal coordinate system about $X_0$ in the 
target manifold. If we write
\bea
	X(z) &=& X_0 + \hbar^{1/2} \xi(z) + O(\hbar) \;,\\
\psiz(z) &=& \psiz_0 + \hbar^{1/2} \eta(z) + O(\hbar)\;, \eea then one
can show that the action~(\ref{susy-action}) may be written
as\footnote{We have implicitly made a change of variables which may
schematically be written as $\eta\to \eta + C\psiz\xi$ to eliminate a
quadratic term of the form $\xi\eta$.} \beq \label{I-sc} I =
i\int_\Sigma dz\wedge d\zbar\;\left[ \left(\dzbar \xi^\mu
\vphantom{\half} \right) \left(\dz\xi^\mu + \half
R^\mu{}_{\sigma\lambda\rho}(X_0) \psiz^\lambda_0 \psiz^\rho_0
\xi^\sigma\right) + \left(\dzbar \eta^\mu\right)\eta^\mu \right] +
O(\hbar^{1/2})\;.  \eeq Note that $\xi$ is a map of $\Sigma$ into
$TM_{X_0}$, \ie, $\xi \in \Map(\Sigma, TM_{X_0})$, and $\eta$ is a
section of $K^{1/2}\otimes X^*_0(TM)$.  $R^\mu{}_{\sigma\lambda\rho}$
is the curvature tensor of the full connection~(\ref{Gamma-full}). 
The steepest descent approximation requires integration over the
normal bundle of the critical point set.  In our case, the (lowest
action) bosonic critical point set is $M$ and we have to integrate
over the normal bundle $NM$ of $M$ in $\mapsm$.  The fibers of $NM$
are spanned by the orthogonal complement to the constant
map\footnote{``Map'' refers to $\xi$ in this discussion.} in
$\Map(\Sigma, TM_{X_0})$.  Integration over the constant maps
corresponds to integrating along $M$.

One last observation is that at $\hbar=0$, equation~ (\ref{I-sc})
defines a free conformal field theory with different Virasoro central
charges for the holomorphic and anti-holomorphic sectors.  This theory
has a conformal anomaly under a Weyl rescaling of the Riemann surface
metric $\gamma_{ab}$.  The change of the path integral under such a
transformation is known and given by the Liouville lagrangian.  For
this reason we choose the metric on $\Sigma$ to be a constant
curvature metric.

By using standard physics methods one can show that the bosonic part
of the measure at $X_0 \in \mapsm$ reexpressed in terms of normal
bundle data is given by
\beq
[\cD X] = \left({\vol\Sigma \over 2\pi}\right)^{d/2} 
\left(d^d X_0\right) [\cD \xi]'\;,
\eeq
where $d = 2n = \dim M$ and $[\cD \xi]'$ is the measure on the space
of maps orthogonal to the constant map. The $\vol\Sigma$ factors
arise because the normalized constant map is $1/\sqrt{\vol\Sigma}$.
The $2\pi$ factors arise from the basic gaussian integral $\int
\exp(-x^2/2) dx = \sqrt{2\pi}\,$. The $d^d X_0$ term means
integration along the critical point set $M$.

Similarly, if $S_0 \sqrt{dz} \in \lamhz(\Sigma)$ is a normalized 
holomorphic spinor \beq \int_\Sigma d^2 z \sqrt{\det \gamma}\; 
(\gamma^{z\zbar})^{1/2} |S_0|^2 =1\;.  \eeq One can always choose 
$\psiz_0^\mu = \eta^\mu_0 \otimes S_0$ and in this way one can 
conclude that 
\beq [\cD\psi] 
= (d^d \eta_0) [\cD \eta]'
\eeq 
where the prime now denotes the 
sections orthogonal to the holomorphic ones. The normalized spinor is 
determined up to an arbitrary phase reflecting the fact that the 
partition function is morally a section of a line bundle.

Let $I_p$ is the identity transformation on a $p$-dimensional vector 
space. The semiclassical approximation gaussian integral can be 
explicitly performed obtaining
\bea
\Zsc(g,\gamma,s) &=& \int (d^{2n} X_0)(d^{2n}\eta_0)\;
\left({\vol\Sigma\over 2\pi}\right)^n \pf'(\dirac\otimes I_{2n})
	\nonumber\\
 &\times &\left[\det{}'\left\{ -(\dzbar\otimes I_{2n})
 \left(\dz \otimes I_{2n} 
+ \half R^\bullet{}_{\bullet\mu\nu} \eta_0^\mu\eta_0^\nu 
S_0^2\right)\right\}\right]^{-1/2}\;.
\eea 
The primes denote that we are working on the space orthogonal to
either the constant maps or the holomorphic spinors, respectively. In
the last line we have been a bit schematic because we want to simplify
the above before writing it in a more intrinsic form. Observe 
that\footnote{To define $\det'\dirac$ we need $\dim\ker\dirac =1$. In 
Section~\ref{final-section} we show that $\Zsc=0$ when $\dim\ker\dirac 
>1$. \label{foot:Zsc}}
$\pf'(\dirac\otimes I_{2n}) = \det'(\dirac\otimes I_n) =
(\det'\dirac)^n$. The rules of Grassmann integration tell us that the
non-vanishing terms must be homogeneous of degree $2n$ in $\eta_0$ and
contain the totally skew expression
$\eta_0^1\eta_0^2\cdots\eta_0^{2n}$. This observation and the fact
that we have to integrate over $M$ allows us to rewrite $\Zsc$ in
terms of integration over differential forms. We define the Cartan
curvature two form by
\beq
	\cR^\mu{}_\nu = \half R^\mu{}_{\nu\rho\sigma}(X_0)\,
	 dX_0^\rho \wedge dX_0^\sigma
\eeq
and we formally think of $\cR$ as a skew linear transformation.
Define an operator $D: \lamzz(\Sigma,TM_{X_0}) \to
\lamzo(\Sigma,TM_{X_0})$ by
\beq
D = \dbar\otimes I_{2n} + \azo\;,
\eeq 
where
\beq
\azo = A_{\zbar} d\zbar = \Szbar^2 d\zbar \otimes \frac{1}{2\pi}\cR\;.
\eeq
We can formally treat $\azo$ as a
flat $SO(2n)$ connection of type $(0,1)$ on the bundle
$\lamzz(\Sigma,X_0^*(TM))$. This allows us to write the semiclassical
partition function as
\beq \label{Z-sc}
\Zsc(g,\gamma,s) = (\vol\Sigma)^n (\det{}'\dirac)^n \;
\int_M \left[ \det{}' D^*(\scalar\otimes I_{2n})\right]^{-1/2} \;,
\eeq 
where $\scalar:\lamzz(\Sigma)\to \lamzo(\Sigma)$. The $(2\pi)$'s
reappear because the only terms of the integrand which contribute are
those which are homogeneous of degree $n$ in $\cR/(2\pi)$. This is our
basic formula; we now compute the determinant.

For reasons which will become clear later on it is convenient to 
change the orientation on $\Sigma$. This interchanges $z$ with $\zbar$.
We do this from now on.

\section{Differential equation for the determinant}
\label{sec-diff}

The parameter space for our determinant is $Y = \Hzo$.  In $\Hzo$ 
there is an integral lattice $L_\Omega$ determined by the period 
matrix $\Omega$ of the Riemann surface $\Sigma$.  The jacobian 
is the complex torus $J(\Sigma)= \Hzo/L_\Omega$.  Basic algebraic
geometric facts about the jacobian are discussed in
Appendix~\ref{app-Q}. 
We are interested 
in the following operators:
\begin{eqnarray}
	\partial & : & \Lzz \to \Loz
	  \\
	D =  *i(\db + \Azo) & : & \Loz \to \Lzz
	  \\
	\BBox_- = D\partial & : & \Lzz \to \Lzz
	  \\
	\BBox_+ = \partial D & : & \Loz \to \Loz
	  \\
	\Delta_- = D D^* & : & \Lzz \to \Lzz
	  \\
	\Delta_+ = D^* D & : & \Loz \to \Loz
\end{eqnarray}
The operator $D$ has index $g-1$.  
When $\Azo$ is a lattice 
point, $D$ is gauge equivalent to $\dbar$ and $\dim\ker D = g$.  If 
$\Azo$ is not a lattice point then $D$ is surjective and $\dim\ker D = 
g-1$.  The operator $\Delta_-$ is invertible if $\Azo$ is not a 
lattice point.  The operator $D$ varies holomorphically on $Y=\Hzo$ 
and due to its equivariance properties it defines a family over the 
jacobian $J(\Sigma)$.

Let $\one$ denote the linear subspace in $\Lzz$ spanned by the 
constant functions and denote its orthogonal complement by $\onep$.  
We remind the reader that $\partial\Azo=0$ and thus 
conclude that $\BBox_- : \onep \to \onep$.  At $\Azo=0$ one 
has that $\BBox_-(\Azo) = \Delta_-$.  Since 
$\Delta_-: \onep \to \onep$ is invertible we have that near 
$\Azo=0$, the operator $\BBox_- : \onep \to \onep$ is 
invertible.  In fact we expect $\BBox_- : \onep \to \onep$ to 
be generically invertible.  The function $\detonep(\Azo)$ is a 
holomorphic function on $\Hzo$ which is nonvanishing near $\Azo=0$.  
We expect the zero set of $\detonep(\Azo)$ to be generically of 
codimension~1 in $\Hzo$.

We are interested in computing $\detonep(\Azo)$ a term in $\Zsc$. 
Since $\det\BBox_{-}$ is not invariant under $L_{\Omega}$, we cannot
evaluate it by elliptic/geometric methods\footnote{The determinant
section of the index zero Dirac operator may be determined using such
methods, see for example \cite{Alvarez-Gaume:1986es}.} on $J(\Sigma)$. 
We note, however, that if we were in finite dimension and all the
operators were invertible $\dy \log\det\BBox_- = \Tr D^{-1}(\dy D)=\dy
\log\det\Delta_-$.  In this way we can relate the derivative $\dy
\log\det\BBox_-$ of a holomorphic function to the geometry of the
determinant line bundle $\DET D \to J(\Sigma)$ because $\det\Delta_-$
gives the Quillen metric on $\DET D$ as we show later.  This
observation is the motivation for the differential equation.

We define the determinant via the heat kernel 
definition
\begin{eqnarray}
	\log\detonep(\Azo) &=& - \int_{\epsilon}^\infty \frac{dt}{t}\;
	\Tronep \e{-t\BBox_-} \;, \\
	&=& - \int_{\epsilon}^\infty \frac{dt}{t}\;
	\left(\Tr \e{-t\BBox_-} - 1\right)\;.
	\label{def-detonep}
\end{eqnarray}
In the above $\epsilon$ is a regularization parameter which we keep 
and at the end of the day we will take the limit $\epsilon \to 
0$ or equivalently 
$$
\log\detonep = 
-\left.\frac{d}{ds}\right|_{s=0}\frac{1}{\Gamma(s)}\int_{0}^{\infty} 
t^{s-1} \Tr \e{-t\BBox_{-}}dt.  
$$
Note that $\detonep$ is a holomorphic function of $\Azo$.  A 
straightforward calculation shows that
\begin{equation}
	\dy\log\detonep(\Azo) = \int_\epsilon^\infty dt \;
	\Tr \e{-t\BBox_+} \partial(\dy D) \;.
	\label{der1}
\end{equation}

We wish to define $\binv$ on $\Lzz$.  We note that $\BBox_- : 
\onep  \to \onep$ is invertible so we know how to define 
$\binv$ on $\onep$. We extend $\binv$ to all of 
$\Lzz$ by defining $\binv(1)=0$. Let $\Pi_{\one}$ be the orthogonal 
projector onto $\one$. We define a partial inverse $D_R^{-1}$ to $D$ by
$$
	D_R^{-1} = \partial \binv\;.
$$
A quick computation shows that
$$
	D D_R^{-1} = \left\{
	\begin{array}{cl}
		I & \quad\mbox{on $\onep$}  \\
		0 & \quad\mbox{on $\one$}\;,
	\end{array}
	\right.
$$
thus concluding that $D D_R^{-1} = I - \Pi_{\one}$. Let $Q = D_R^{-1} D = 
\partial \binv D$. Note that the image of $Q$ is contained in 
$\Hoz^{\perp}$. One verifies that $Q^2=Q$ and thus $Q$ is a 
projector but not an orthogonal projector. To characterize $Q$ we 
observe the following. 

If $\Azo$ is not a lattice point then $D$ is surjective and there 
exists $\eta\in\Loz$ such that $D\eta=1$ and $Q\eta=0$.  
We note that $Q: \Hoz^{\perp}\to\Hoz^{\perp}$.  This may be seen 
by observing that if $\omega\in \Hoz^{\perp}$ then $\exists f 
\in\Lzz$ such that $\partial f = \omega$.  Using the definition of $Q$ 
one immediately sees that $Q\omega=\omega$.  Also, if $\phi\in 
\ker D$ then $Q\phi=0$.  This allows us to conclude that $\ker 
\BBox_+ = \ker D \oplus \bbC\eta$. The projector $Q:\Loz \to \Loz$ may be 
characterized by
$$
	Q = \left\{
	\begin{array}{cl}
		I & \quad\mbox{on}\; \Hoz^{\perp}  \\
		0 & \quad\mbox{on}\; \ker\BBox_+
	\end{array}
	\right.
$$

We now return to (\ref{der1}) and insert the decomposition $I = 
DD_R^{-1} + \Pi_{\one}$.
\begin{eqnarray}
	\dy\log\detonep &=& \int_\epsilon^\infty dt\;
	\Tr \e{-t\BBox_+}\partial(\dy D) 
		\nonumber\\
	&=& \int_\epsilon^\infty dt\;
	\Tr \e{-t\BBox_+}\partial(D\partial\binv +\Pi_{\one})(\dy D) 
		\label{der2}\\
	&=& \int_\epsilon^\infty dt\;
	\Tr \e{-t\BBox_+}\BBox_+\partial\binv(\dy D) 
		\nonumber \\
	&=& - \int_\epsilon^\infty dt\; \frac{d}{dt}
	\Tr \e{-t\BBox_+} \partial\binv(\dy D)
		\nonumber\\
	&=& - \left. \Tr \e{-t\BBox_+} \partial\binv(\dy D) 
	\right|_\epsilon^\infty\;.
		\label{der3}
\end{eqnarray}
In (\ref{der2}) we used $\partial\Pi_{\one}=0$.

First we investigate $\lim_{t\to\infty} \e{-t\BBox_+}$.  We denote the
limiting operator by $\e{-\infty\BBox_+}$.  Assume $\psi$ is an
eigenvector of $\BBox_+$ with nonzero eigenvalue $\lambda$.  Thus we
can write $\psi = \lambda^{-1}\BBox_+ \psi = \lambda^{-1}\partial
D\psi$.  Because of the $\partial$ we see that $\psi\in \Hoz^{\perp}$. 
We have a non-orthogonal decomposition $\Loz = \Hoz^{\perp} \oplus
\ker\BBox_+$.  It is easy to see  that for $\Azo$ sufficiently near zero
that $\lambda$ has a positive real part.  From now on we assume that
$\Azo \in \tilde{V}$ where $\tilde{V}$ is a small neighborhood of $0$
with $0$ deleted.  We see that $\e{-\infty\BBox_+}(\Hoz^{\perp})= 0$. 
Also $\e{-\infty\BBox_+} = I$ on $\ker\BBox_+$.  Thus we conclude that
$\e{-\infty\BBox_+} = I-Q$.  We now go back to (\ref{der3}) and write
\begin{eqnarray}
	\dy\log\detonep &=& - \Tr \e{-\infty\BBox_+} \partial \binv(\dy D)
	\nonumber \\
	&+& \Tr \e{-\epsilon \BBox_+} \partial\binv(\dy D) \;.
	\nonumber
\end{eqnarray}
In the first line we observe that $\e{-\infty\BBox_+} \partial=0$ 
because $\partial$ maps onto $\Hoz^{\perp}$ and $\Tr 
\e{-\infty\BBox_+}$ annihilates $\Hoz^{\perp}$.  Thus we conclude that
\begin{equation}
	\dy\log\detonep = \Tr \e{-\epsilon \BBox_+} \partial\binv(\dy D)\;.
	\label{der-det}
\end{equation}
In a similar fashion we can also show that
\begin{equation}
	\dy\log\det\Delta_- 
	= \Tr \e{-\epsilon \Delta_+} D^*\Delta_-^{-1}(\dy D)\;.
	\label{der-det-lap}
\end{equation}
Note that the right hand side of the two last equations involves
inverses to $D$.  The regularization and the infinite dimensional
nature of the problem lead to slight differences between the right
hand sides.  We now compute those differences.  We first observe that
\begin{eqnarray}
	\dy\log\detonep - \dy\log\det\Delta_- 
	& = & 
		 \Tr\left( \e{-\epsilon \BBox_+} - \e{-\epsilon \Delta_+} 
		 \right) D^*\Delta_-^{-1}(\dy D)
	\nonumber \\
	 & + & 
		\Tr \e{-\epsilon \BBox_+}(\partial\binv - D^*\Delta_-^{-1})
		(\dy D)  
	 \label{eq3}
\end{eqnarray}

The first line of (\ref{eq3}) is explicitly computable.  The 
computation is a long and tedious exercise in perturbation theory.  A 
key observation is that $D^{-1}(z,z') \sim (z-z')^{-1}$ as $|z-z'|\to 
0$.  We state the result  as a proposition
\begin{proposition}
$$ \lim_{\epsilon\to 0} \Tr\left( \e{-\epsilon \BBox_+} - \e{-\epsilon \Delta_+} 
	 \right) D^*\Delta_-^{-1}(\dy D) =
	 - \frac{i}{2\pi} \int_\Sigma \Azob \wedge \dy \Azo\;.
$$
\end{proposition} 

The operator $(\partial\binv - D^*\Delta_-^{-1})$ appearing in the 
second line of (\ref{eq3}) is of finite rank.  The basic reason is 
that $\partial\binv$ and $D^*\Delta_-^{-1}$ are both partial inverses 
to $D$.  The easiest way to demonstrate finite rank is to observe that 
$Q(\partial\binv)=\partial\binv$, $Q(D^*\Delta_-^{-1})=\partial\binv$ 
and therefore the image of $(\partial\binv-D^*\Delta_-^{-1})$ must be in 
$\ker Q$.  The finite rank condition allows us to safely take the 
$\epsilon\to 0$ limit in the second line of (\ref{eq3}).  We 
explicitly take  $\epsilon\to 0$ and obtain
\begin{eqnarray}
	\dy\log\detonep - \dy\log\det\Delta_- 
	& = & 
	 	-\frac{i}{2\pi} \int_\Sigma \Azob \wedge \dy \Azo\;.
	\nonumber\\
	 & + &  
		\Tr \left(\partial\binv - D^*\Delta_-^{-1}\right)(\dy D)
	 \label{der4}
\end{eqnarray}

Let $\{\omega_i\}_{i=1}^g$ be 
a standard basis for $\Hoz$.  First we choose a holomorphically 
varying basis $\{\tau_a\}_{a=1}^{g-1}$ for $\ker D$ over 
$\widetilde{V}$.  Next, let $\{\xi_\alpha\}_{\alpha=1}^g$ be a 
holomorphically varying basis over $\widetilde{V}$ for $\ker\BBox_+$ 
chosen as $\{\tau_a\}_{a=1}^{g-1} \cup \{\eta\}$,  where $\eta$ is a 
solution to $D\eta=i$ which varies holomorphically over 
$\widetilde{V}$. It is convenient to define two matrices, the $g 
\times g$ matrix
\begin{equation}
	M_{i\alpha} = \langle \omega_i, \xi_\alpha\rangle 
	\label{def-M}
\end{equation}
and the $(g-1)\times(g-1)$ matrix 
\begin{equation}
	h_{ab} = \langle\tau_a, \tau_b\rangle\;.
	\label{def-h}
\end{equation}
We use the physics convention that the hermitian inner product 
$\langle\cdot,\cdot\rangle$ is anti-linear in the first slot.
\begin{proposition}
$
	\Tr \left(\partial\binv - D^*\Delta_-^{-1}\right)(\dy D) =
	\dy \log\det M - \dy \log \det h\;.
$
\end{proposition}
The proof is a straightforward use of the variational formulas of 
perturbation theory.

Going back to (\ref{der4}) we see that the combination 
$\det\Delta_-/\det h$ appears. The first remark is that
$$
	\det\Delta_- = \det_{(\ker D)^{\perp}} \Delta_+\;.
$$
Thus the combination
\begin{equation}
	\frac{1}{\det h}\; \det_{(\ker D)^{\perp}} \Delta_+
	\label{Q-metric}
\end{equation}
appears in (\ref{der4}).  This is important because the hermitian 
Quillen metric for the determinant line bundle $\DET D \to J(\Sigma)$ 
in the trivialization given by 
$(\tau_1\wedge\cdots\wedge\tau_{g-1})^{-1}$ is precisely 
(\ref{Q-metric}).  The Quillen connection in this trivialization is 
given by the one-form
\begin{equation}
	\nutil' = \dy \log\left(\frac{1}{\det h}\; 
	\det_{(\ker D)^{\perp}} \Delta_+ \right)
	\label{Q-conn}
\end{equation}
Define a one-form on $\Hoz$ by
\begin{equation}
	\rho = \frac{i}{2\pi} \int_\Sigma \overline{\Azo}\wedge d_Y\Azo\;.
	\label{def-rho}
\end{equation}
One can show that
\begin{equation}
	d\nutil'= \dyb\nutil'=d\rho=\dyb\rho = \frac{i}{2\pi}\; \int_\Sigma
		d_Y \Azob \wedge d_Y \Azo \;.
	\label{Q-curv}
\end{equation}
One can interpret $d\nutil'$ as the curvature of the Quillen connection. 
It is the standard translationally invariant symplectic $2$-form    
on the jacobian, the polarization.

It is illustrative to see the above in genus~$1$.  The
complex coordinate on a torus $\Sigma$ with modular parameter $\tau = 
\tau_1 + i\tau_2$ will be denoted by $\zeta$.  The jacobian 
$J(\Sigma)$ and $\Hzo$ have complex coordinate $z$.  We write $\Azo = 
\pi z d\bar{\zeta}/\tau_2$.  If we also write $z = u + \tau v$ then 
the jacobian torus corresponds to period~$1$ in the real coordinates 
$u$ and $v$.  One immediately verifies that the Quillen curvature 
two-form~(\ref{Q-curv}) is $d\nutil = -2\pi i du\wedge dv$.

Using the above definitions and derivations we see that  $\det 
M/\detonep$ satisfies the partial differential equations
\begin{eqnarray}
	\left(\dy+\nutil'-\rho\right)\left(\frac{\det M}{\detonep}\right) 
	&=& 0\;, \label{pde-detM} \\
	\dyb \left(\frac{\det M}{\detonep}\right) & = & 0\;.
	\label{pde-detM1}
\end{eqnarray}
To  evaluate $\det M/\detonep$ we have to 
understand what happens at the origin $\Azo=0$ which we do in 
Section~\ref{detM-section}.

\section{Holomorphic Trivialization; the $\vartheta$-function}
\label{hol-triv}

In this section, we trivialize $\DET D_{A}$ pulled up to $\Hzo$ and 
realize a cross section as a $\vartheta$-function.

\subsection{The flat trivialization}

Let $\cL\to J(\Sigma)$ be the determinant line bundle of $D_A = 
*i(\dbar+\azo): \Loz\to\Lzz$ over the jacobian of a Riemann 
surface~$\Sigma$.  Let $\pi:\Hzo\to J(\Sigma)$ be the standard 
covering map.  We have a pull-back holomorphic line bundle $\cLt = 
\pi^* \cL$ which can be holomorphically trivialized over $\Hzo$.  We 
will now discuss an explicit holomorphic trivialization of $\cLt$.

$\cL$ has its Quillen connection $\nu$.  Let $\nutil = \pi^* 
\nu$ be the pull-back of this connection.  Let $\rho$ 
(see~(\ref{def-rho})) be the 1-form on $\Hzo$ given by
$$
	\rho = \frac{i}{2\pi}\int_\Sigma \Azob \wedge d_{\Hzo} \azo\;.
$$
We have shown that $d(\nutil-\rho)=0$ so that $\nutil-\rho$ is a flat 
connection on $\cLt$.  Trivialize $\cLt$ by parallel transport from 
$A$ to $0$ via this flat connection along any path.  Since 
$\dbar(\nutil-\rho)=0$ the trivialization is holomorphic.  If $\cLt_0$ 
is the fiber over $0$, this trivialization, which we call the {\em 
flat trivialization\/}, is given by a map $\varphi: \cLt \to 
\Hzo\times\cLt_0$ defined by the parallel transport map
\begin{equation}
	\tilde{s} = \exp\left(-\int_0^A(\nutil-\rho)\right)
	(\varphi\tilde{s})\;,
	\label{s-par}
\end{equation}
where $\tilde{s} \in \cLt_A$.

If $s$ is a section of $\cL$ then the pull-back section 
$\tilde{s}=\pi^* s$ satisfies the periodicity requirement 
\begin{equation}
	\tilde{s}(A+B) =\tilde{s}(A)
	\label{s-per}
\end{equation}
where $B$ is a lattice vector. Using the 
definition of $\varphi$ one can now show that
\begin{eqnarray}
	(\varphi\tilde{s})(A+B) & = & e^{\int_0^B \nutil}  \nonumber\\
	 & \times & e^{\frac{i}{4\pi} \int_\Sigma \Bzob\wedge\Bzo}
	 e^{\frac{i}{2\pi}\int_\Sigma \Bzob\wedge\azo}
	 (\varphi\tilde{s})(A)\;. \label{flat-cocycle}
\end{eqnarray}
In the above $e^{\int_0^B \nutil}$ is along the straight line from $0$
to $B$.  Note that the cocycle defined by
$(\varphi\tilde{s})(A+B)/(\varphi\tilde{s})(A)$ is holomorphic because
it does not depend on $\Azob$.  To prove (\ref{flat-cocycle}) insert
(\ref{s-par}) into (\ref{s-per}) with the result that
$$
(\varphi\tilde{s})(A+B) = e^{\int_0^{A+B} (\nutil-\rho)} 
	e^{\int_0^A (\nutil-\rho)} (\varphi\tilde{s})(A)
$$
where all integrals are along straight lines joining the respective 
points. Using the flatness of $\nutil-\rho$ the above may be written as
$$
(\varphi\tilde{s})(A+B) = e^{\int_A^{A+B} (\nutil-\rho)} 
(\varphi\tilde{s})(A)	
$$
Using the definition~(\ref{def-rho}) of $\rho$ we have
$$
	\int_A^{A+B} \rho = 
	- \frac{i}{2\pi}\; \int_\Sigma \Azob\wedge\Bzo 
	-\frac{i}{4\pi}\; \int_\Sigma \Bzob\wedge\Bzo\;.
$$
We observe that
$$
	\int_0^A \nutil + \int_A^{A+B} \nutil 
	- \int_B^{A+B} \nutil - \int_0^B \nutil = \int_P d\nutil\;,
$$
where $P$ is the parallelogram spanned by $A$ and $B$. Note that
$$
	\int_0^A \nutil - \int_B^{A+B} \nutil =
	\int_0^A \pi^*\nu - \int_B^{A+B} \pi^*\nu = 0\;,
$$
because $B$ is a lattice vector and $\nutil$ is the pullback of 
$\nu$. Thus we conclude
$$
	\int_A^{A+B} \nutil 
	= \int_0^B \nutil + \int_P d\nutil\;.
$$
We see that
	\begin{equation}
		e^{-\int_0^B \nutil} = e^{-\int_0^B \pi^*\nu}
		= e^{-\int_{\gamma_B} \nu} = \hol(\gamma_B)\;,
		\label{holonomy}
	\end{equation}
where the loop $\gamma_B \subset J(\Sigma)$ is the projection on the 
jacobian of the straight line from $0$ to $B$.  $\hol(\gamma_B)$ 
is the holonomy of the Quillen connection.  The remaining computation 
using (\ref{Q-curv}) gives
$$
	\int_P d\nutil = \frac{i}{2\pi} \int_\Sigma \Bzob\wedge\Azo
	- \frac{i}{2\pi} \int_\Sigma \Azob\wedge\Bzo \;.
$$
Putting together the various terms we obtain~(\ref{flat-cocycle}).

We can be more explicit by writing $A$ and $B$ as
\begin{eqnarray}
	\azo & = & 2\pi i \sum 
	z_j(\Omega-\bar{\Omega})^{-1}_{jk}\bar{\omega}_k\;,  
	\label{z-coord}\\
	\Bzo_{nm} & = & 2\pi i \sum (m + \Omega n)_j 
	(\Omega-\bar{\Omega})^{-1}_{jk} \bar{\omega}_k\;.
	\label{B-coord}
\end{eqnarray}
The $(z_1, \ldots, z_g)$ are the standard complex coordinates on 
$\Hzo$.  We will often abbreviate $\Bzo_{nm}$ to $\Bzo$.  One can now 
verify that
\begin{eqnarray*}
	\frac{i}{4\pi}\int_\Sigma \Bzob\wedge\Bzo & = & 
	i\pi\sum_{j,k}(m+\bar{\Omega}n)_j  
	(\Omega-\bar{\Omega})^{-1}_{jk} (m +\Omega n)_k\;,\\
	\frac{i}{2\pi}\int_\Sigma \Bzob\wedge\azo & = & 
	2\pi i \sum_{j,k}(m + \bar{\Omega} n)_j
	(\Omega-\bar{\Omega})^{-1}_{jk} z_k \;.
\end{eqnarray*}
By inserting the above into 
(\ref{flat-cocycle}) one sees that this is not the standard cocycle 
that defines a theta function.

\subsection{The standard trivialization}

To get the {\em standard trivialization} $\Phi$ we multiply $\varphi$ by a 
specific nonvanishing holomorphic function:
\begin{equation}
	(\Phi\tilde{s})(A) = \exp\left(-\pi i \sum_{j,k} z_j
	(\Omega-\bar{\Omega})^{-1}_{jk} z_k\right)(\varphi\tilde{s})(A)\;.
	\label{std-cocycle}
\end{equation}
After some algebra one finds
\begin{equation}
	(\Phi\tilde{s})(A+B)  =  \chi(B_{nm})
	  \times  e^{-\pi i \sum_{j,k} n_j \Omega_{jk} n_k}
	 e^{-2\pi i \sum n_j z_j} (\Phi\tilde{s})(A)\;.
	\label{theta-cocycle}
\end{equation}
The above is the standard transformation law for a theta function 
where
\begin{eqnarray}
	\chi(B_{nm}) &=& e^{-\pi i \sum m_j n_j}e^{\int_0^B \nu}\\
	& = & e^{-\pi i \sum m_j n_j} \hol(\gamma^0_{nm})^{-1} \;.
	\label{def-chi}
\end{eqnarray}
is a character for the lattice.  The closed curve $\gamma^0_{nm}$ is 
the projection into the jacobian of the straight line from $0$ to 
$B_{nm}$ in $\Hzo$.  Verifying that $\chi$ is a character is based on 
the following identity. Given two lattice vectors $B$ and $B'$
$$
	\int_0^B \nutil + \int_B^{B+B'} \nutil- \int_0^{B+B'} \nutil=
	\int_\triangle d\nutil\;,
$$
where $\triangle$ is the triangle with vertices $\{0, B, B+B'\}$.
A quick computation shows that
$$
	\int_\triangle d\nutil = \frac{i}{4\pi}\;
	\int_\Sigma\left(\Bzobp\wedge\Bzo - \Bzob\wedge\Bzop\right)\;.
$$
Using (\ref{B-coord}) and an obvious notation we see that
$$
	\int_\triangle d\nutil = i\pi(m'\cdot n - m \cdot n')\;.
$$
Remembering that $B$ and $B'$ are lattice vectors we have 
$\int_B^{B+B'}\nutil = \int_0^{B'}\nutil$. We conclude
$$
	\int_0^B \nutil + \int_0^{B'} \nutil - \int_0^{B+B'} \nutil =
	i\pi(m'\cdot n - m \cdot n')\;.
$$
It is now straightforward verifying that (\ref{def-chi}) is a 
character for the lattice.
If one writes
\begin{equation}
	\chi(B_{nm}) = e^{-2\pi i n\cdot b}e^{2\pi i m\cdot a}
	\label{def-char}
\end{equation}
then (\ref{theta-cocycle}) is the transformation law for the theta 
function $\vartheta\left[{a \atop b}\right](z)$.

\subsection{Algebraic geometry viewpoint}
\label{sec-alg-geom}

The holomorphic line bundle $\cL$ depends on the basepoint $P_0$. We 
have exhibited a cross section $\vartheta\left[{a \atop b}\right](z)$ 
of $\pi^*(\cL)$ by trivializing $\pi^*(\cL)$ using differential 
geometry. In Appendix~\ref{app-kappa}, we summarize the algebraic 
geometric construction of $\vartheta\left[{a \atop b}\right]$ which 
also tells us that $\left[{a \atop b}\right]$ in $J_0(\Sigma)$ 
equals the Riemann constant $-\kappa$.
Thus $\vartheta\left[{a \atop b}\right]$ is the translate of the 
ordinary theta function $\vartheta\left[{0 \atop 0}\right]$ by 
$-\kappa$; an explicit formula for $\kappa$ can be found 
in \cite{Mumford:1}.

\section{The explicit construction of $\DET D\to J(\Sigma)$}
\label{detd-construction}

The construction of the determinant line bundle is particularly simple 
in this case because the kernel jumps only at the origin of $J(\Sigma)$. 
We go through the details because we need an explicit representation 
of transition functions.

Let $O$ be the origin in $J(\Sigma)$.  Let $V = J(\Sigma) - \{O\}$ and 
let $U$ be a small neighborhood of $O$.  On $V$ we have that $\dim\ker 
D=g-1$ and $\ker D$ is a rank $g-1$ vector bundle over $V$.  We can 
take its determinant and obtain the determinant line bundle $\cL_V \to 
V$.  On $U$ we use the operator $\cDt: \Loz \to \Lzz$ defined by
\begin{equation}
	\cDt\phi = D\phi - i\int_\Sigma \left(\Azo\wedge\phi\right) \;
\end{equation}
for $\phi \in \Loz$.  A little thought shows that for $\Azo\in U$, 
$\dim\ker\cDt=g$ and we can define a determinant line bundle $\cL'_U$ 
over $U$ for the operator $\cDt$.  We will patch these bundles to 
construct the bundle $\cL \to J(\Sigma)$.

We observe that $\ker \cDt$ on $U\cap V$ contains $\ker D$ as a 
subspace of codimension~$1$.  On $U\cap V$ we have the exact sequence 
of vector bundles
$$
	0 \to \ker D \to \ker \cDt \to \mathcal{K} \to 0\;,
$$
where $\mathcal{K} = \ker\cDt/\ker D$.  As a consequence $\cL'_U$ is 
isomorphic to $\cL_V \otimes \mathcal{K}$ over $U\cap V$.  To patch 
$\cL'_U$ with $\cL_V$ we have to locally trivialize $\mathcal{K}$ 
which we now do.  We first identify $U$ with a deleted neighborhood of 
the origin in $\Hzo$ with coordinates $(z_1,\ldots,z_g)$ given by 
(\ref{z-coord}).  We then introduce an open cover $\{V_1,\ldots,V_g\}$ 
of $U \cap V$, where the open set $V_j$ is the set in $U\cap V$ with 
$z_j\neq 0$.  On $V_j$ we can find a holomorphically varying basis 
$\{\tau_1^{(j)},\ldots,\tau_{g-1}^{(j)}\}$ for $\ker D$ and a 
holomorphically varying $\eta^{(j)}$ satisfying 
\begin{equation}
	D\eta^{(j)}=i\;.
	\label{def-eta}
\end{equation}
As a result of the last equation, $D(\eta^{(j)}-\eta^{(k)})=0$ on 
$V_j\cap V_k$, so $\eta^{(j)} \cong \eta^{(k)} \pmod{\ker D}$.  
Therefore, the equivalence class of $\{\eta^{(j)}\}_{V_j}$ gives a 
holomorphic section of $\mathcal{K}$.  This section gives the 
isomorphism of $\cL_V$ with $\cL'_U$ over $U\cap V$.  Patching 
together $\cL_V$ with $\cL'_U$ gives us the determinant line bundle 
$\DET D \to J(\Sigma)$.

To get explicit transitions functions on $V_j\cap V_k$, we need 
an explicit holomorphic basis for $\ker D$.  A generic element 
$\tau$ in $\ker D$ satisfies the equation $\dbar \tau + 
\azo\wedge\tau=0$ from which it follows that $\int_\Sigma 
\azo\wedge\tau=0$.  The differential equation is equivalent to the 
integral equation
\begin{equation}
	\tau = \omega - 
	\partial\Delta_0^{-1}\left[*(\azo\wedge\tau)\right]\;,
	\label{tau-int}
\end{equation}
where $\omega \in \Hoz$ and $\int_\Sigma \azo\wedge\omega=0$. On the 
other hand, $\eta$ satisfies the equation $*(\dbar\eta + 
\azo\wedge\eta) = 1$ from which follows that $\int_\Sigma 
(\azo\wedge\eta) = 1$; we have normalized the metric such that 
the volume of $\Sigma$ is one. The associated integral equation is
\begin{equation}
	\eta = \omega - 
	\partial\Delta_0^{-1}\left[*(\azo\wedge\eta)-1\right]\;,
	\label{eta-int}
\end{equation}
where $\omega \in \Hoz$ and $\int_\Sigma \azo\wedge\omega=1$.  We 
can now explicitly trivialize the line bundles $\cL'_U$ and 
$\cL_V$ in $V_j$.  We order the $\tau$'s and $\eta$ in a specific 
way to simplify signs.  Let $(\xi^{(j)}_1,\ldots,\xi^{(j)}_g) = 
(\tau^{(j)}_1, \tau^{(j)}_2,\ldots, \tau^{(j)}_{j-1}, \eta^{(j)}, 
\tau^{(j)}_j,\ldots,\tau^{(j)}_{g-1})$.  For $k\neq j$ we define 
$\xi^{(j)}_k$ by choosing $\omega$ in eq~(\ref{tau-int}) to be 
$\omega_k -(z_k/z_j)\omega_j$.  We define $\eta^{(j)} = 
\xi^{(j)}_j$ by choosing $\omega$ in eq~(\ref{eta-int}) to be 
$-\omega_j/z_j$.  Note that if $U$ is chosen small enough then 
integral equations (\ref{tau-int}) and (\ref{eta-int}) imply that 
the $\xi^{(j)}$'s vary holomorphically over $V_j$.

We trivialize $\cL_V$ in $V_j$ via the local holomorphic section
$$
	(-1)^{g-(j-1)} \;
	{1 \over \tau^{(j)}_1 \wedge\cdots\wedge \tau^{(j)}_{g-1}}
$$
which gives the trivialization
$$
	(-1)^{g-(j-1)} \;
	{1 \over \tau^{(j)}_1 \wedge\cdots\wedge 
	\tau^{(j)}_{g-1}\wedge\eta^{(j)}}
$$
of $\cL'_U$ in $V_j$.  If we use the integral equation one can 
check that the transition function for $\cL_V$ to go from $V_j$ to 
$V_k$ is given by $z_j/z_k$. These transition functions are familiar 
from the study of the canonical line bundle on $\CP^n$.

\relax
\section{Evaluation of det $\Box_{-}$}
\label{detM-section}

Let $\xi \in \Loz$ and define
$$
	P\xi = *(\azo\wedge\xi) - \int_\Sigma(\azo\wedge\xi)\;.
$$
On $U$ we have an isomorphism with between $\Hoz$ and $\ker\cDt$ 
defined by the integral equation
\begin{equation}
	\varpi = \omega - \partial\Delta_0^{-1}P\varpi\;,
	\label{pdevarpi}
\end{equation}
where $\omega\in\Hoz$ and $\varpi\in\ker\cDt$.  Let 
$\varpi_j$ be the image of $\omega_j$.  A 
consequence of the integral equation is that 
$\{\varpi_1,\ldots,\varpi_g\}$ varies holomorphically on $U$.  On $U$, 
we can identify the line bundle $\cL'_U$ with the line bundle $U\times 
(\bigwedge^{\rm top}\Hoz)^{-1}$ thus trivializing $\cL'_U$.  Denote 
the isomorphism of the dual bundles $(\cL'_U)^*$ with $U\times 
(\bigwedge^{\rm top}\Hoz)$ by $f$.  Now define a section $\mu$ of 
$\cL_V$.  If $y$ is a point in the fiber of $(\cL_V)^*$ over $\azo$ 
then $y \otimes \eta$ is in $(\cL'_U)^*_{\azo}$.  We define $\mu(y) = 
f(y \otimes \eta)$.  $\det M^{(j)}$ is the function 
$\mu(\tau_1^{(j)}\wedge\cdots\wedge\tau_{g-1}^{(j)})$.  This agrees 
with our previous definition of $\det M$ defined in $V_j$.  If 
$\xi=(\tau_1,\ldots,\tau_{g-1},\eta)$ then $M_{i\alpha} = \langle 
\omega_i,\xi_\alpha\rangle$ and $\det M = \det(M_{i\alpha})$.  Said 
differently we have
$$
	\mu = (\det M^{(j)})\;{1\over 
	\tau_1^{(j)}\wedge\cdots\wedge\tau_{g-1}^{(j)}}
$$
over $V_j$.

The trivial bundle $U\times (\bigwedge^{\rm top}\Hoz)^{-1}$ has a 
canonical section
$$
	{1\over \omega_1\wedge\cdots\wedge\omega_g}\;.
$$
A local holomorphic section of $\cL'_U$ is 
\begin{equation}
	{1 \over \varpi_1\wedge\cdots\wedge\varpi_g}\;,
	\label{lp-sec}
\end{equation}
which gives trivialization of $\cL'_U$ which we call the 
$U$-trivialization.
On $V_j$ we had the trivialization of $\cL'_U$ given by
\begin{equation}
	{1 \over 
	\tau^{(j)}_1\wedge\cdots\wedge\tau^{(j)}_{g-1}\wedge\eta^{(j)}}\;.
	\label{lv-sec}
\end{equation}

One can immediately check that the ratio of (\ref{lp-sec}) to 
(\ref{lv-sec}) is given by
$$
	{\det H \over \det M}\;.
$$
Here $H_{ij}$ is the matrix $\langle \omega_{i},\omega_{j}\rangle$.
Hence the section $s_U/\varpi_1\wedge\cdots\wedge\varpi_g$ of $\cLt$ 
is identified with the section 
$s_{V_j}/\tau^{(j)}_1\wedge\cdots\wedge\tau^{(j)}_{g-1}$ of $\cL_V$ 
trivialized over $V_j$ where
	\begin{equation}
		s_U = s_{V_j} \; {\det H \over \det M^{(j)}}\;.
		\label{theta-trans}
	\end{equation}

Since $\mu$ is represented by $\det M^{(j)}$ in the $V_j$ 
trivialization, it is represented by the constant $\det H$ in 
the $U$ trivialization. Thus we see that $\mu$ is a well 
defined holomorphic section of $\cL'_U$ and it does not vanish at the 
origin.

We showed that $\det M^{(j)}$ satisfies the equations
\begin{eqnarray}
	\left[\partial + (\nutil^{(j)}-\rho)\right]
	\left({\det M^{(j)} \over \detonep}\right) &= &0\;, 
	\label{eq1}\\
	\dbar\left({\det M^{(j)} \over \detonep}\right) &= &0\;,
	\label{eq2}
\end{eqnarray}
where $\nutil^{(j)}$ is the one form which represents the 
connection $\omega$ in the $V_j$ trivialization.
Thus ${\mu /\detonep}$ may be viewed as a covariantly constant
holomorphic section of the line bundle $\cL'_U$.  We conclude that
\begin{equation}
	{\mu(A)\over\detonep(A)} = e^{-\int_0^A(\nutil-\rho)} s_1
	\label{mu-pt}
\end{equation}
for some $s_1$ in $\cLt_0$.  We define
the theta section $\theta$ in the standard trivialization by  
\begin{equation}
	\theta(A) = 
	e^{\pi i \sum z_j (\Omega-\bar{\Omega})^{-1}_{jk} z_k}
	\vartheta\left[{a \atop b}\right](A)\;
	e^{-\int_0^A(\nutil-\rho)} s_0\;
	\label{theta-pt}
\end{equation}
where $s_0 \in \cLt_0$.
Taking the ratio of (\ref{theta-pt}) to (\ref{mu-pt}) we have
$$
	\detonep(A)\; {\theta(A) \over \mu(A)} = 
	e^{\pi i \sum z_j (\Omega-\bar{\Omega})^{-1}_{jk} z_k}\;
	\vartheta\left[{a \atop b}\right](A)\;
	{s_0\over s_1}\;.
$$
Note that $s_0/s_1$ is a complex number.
We evaluate the ratio of the two sections on the left hand 
side by choosing an appropriate trivialization.  The ratio is simplest 
in the $U$ trivialization where 
\begin{eqnarray}
	\mu &= & \frac{\det H}{\varpi_1\wedge\cdots\wedge\varpi_g}\;,
		\label{def-mu}\\
	\theta &=& \frac{\zanalog}{\varpi_1\wedge\cdots\wedge\varpi_g}\;.
		\label{def-theta-u}
\end{eqnarray}
Formula (\ref{def-theta-u}) defines $\zanalog$ which is central in 
what follows. The zeroes of $\zanalog$ describe the $\Theta$-divisor 
in a small neighborhood of the origin in the jacobian.
Thus we conclude that
\begin{equation}
	\detonep(A) = c e^{\pi i \sum z_j (\Omega-\bar{\Omega})^{-1}_{jk} z_k}\;
	{\vartheta\left[{a \atop b}\right](A) \over \zanalog(A)}\;,
	\label{detonep}
\end{equation}
where $c$ is a constant to be determined by the behavior at $A=0$. Let
$$
	r = \lim_{A\to 0} 
	{\vartheta\left[{a \atop b}\right](A) \over \zanalog(A)}
$$
then
\begin{equation}
	\detonep(A) = \frac{1}{r}\;
	\detp\; e^{\pi i \sum z_j (\Omega-\bar{\Omega})^{-1}_{jk} z_k}\;
	{\vartheta\left[{a \atop b}\right](A) \over \zanalog(A)}\;,
\end{equation}

There is an interesting expression for the ratio $r$.  The 
$\vartheta$-function in the standard trivialization is given 
by~(\ref{theta-pt}).  On the other hand we know that $\theta = 
\zanalog/\varpi_1\wedge\cdots\wedge\varpi_g$.  Taking ratios and 
limits we conclude that
\begin{equation}
	r = {(\omega_1\wedge\cdots\wedge\omega_g)^{-1} \over s_0}\;.
	\label{def-r}
\end{equation}
We have a choice in the overall scale of the $\theta$-section so we 
can make the constant $r=1$.

\begin{theorem}
	\label{thm-detonep}
\begin{equation}
	\detonep(A) = 
	\detp\; 
	{\vartheta\left[{a \atop b}\right](A) \over \zanalog(A)}\;
	\exp\left(\pi i \sum_{j,k} 
	z_j (\Omega-\bar{\Omega})^{-1}_{jk} z_k\right)\;.
	\label{detonep-final}
\end{equation}
	In the above we have normalized the theta section so that
$$
	\lim_{A\to 0} 
	{\vartheta\left[{a \atop b}\right](A) \over \zanalog(A)} =1\;.
$$
	The characteristics of the theta function are determined by the 
	holonomy of the Quillen connection for the determinant line bundle 
	$\DET D \to J(\Sigma)$ by equations (\ref{def-chi}) and 
	(\ref{def-char}).  They turn out to be $-\kappa$, with $\kappa$ 
	the Riemann constant; see  Section~\ref{sec-alg-geom} and
	Appendix~\ref{app-kappa}.  Although $\kappa$ depends on the 
	choice of basepoint $P_0$, the ratio 
	$\vartheta[-\kappa](A)/\zanalog(A)$ does not.
\end{theorem}
	
\subsection{The determinant in the case of genus $1$}

In this section we show how the previous applies to genus~$1$.  The 
complex coordinate on a torus $\Sigma$ with modular parameter $\tau = 
\tau_1 + i\tau_2$ will be denoted by $\zeta$.  The jacobian 
$J(\Sigma)$ and $\Hzo$ have complex coordinate $z$.  We write $\Azo = 
\pi z d\bar{\zeta}/\tau_2$.  If we also write $z = u + \tau v$ then 
the jacobian torus corresponds to period~$1$ in the real coordinates 
$u$ and $v$.  One immediately verifies that the Quillen curvature 
two-form~(\ref{Q-curv}) is $d\omega = 2\pi  du\wedge dv$.

It is clear from the differential equation~(\ref{def-eta}) that $\eta 
\propto d\zeta/z$ and thus $\det M \propto 1/z$.  The theta line 
bundle over $J(\Sigma)$ has first Chern class equal to~$1$.  There is 
a unique holomorphic section $\theta$ with a simple zero.  Assume the 
zero is at the origin~$O$ in $J(\Sigma)$.  This line bundle is 
characterized by the divisor~$O$.  Choose $U$ and $V$ as in the 
beginning of Section~\ref{detd-construction}.  We can characterize 
this section by saying that $\zanalog=z$ and $\theta_V=1$.  Comparing 
with equation~(\ref{theta-trans}) we see that $z = \zanalog/\theta_V 
\propto 1/\det M$ in agreement with our general results.  Using 
Theorem~\ref{thm-detonep} we conclude that
$$
		\detonep(z) = 
	\detp\; 
	{\vartheta\left[{1/2 \atop 1/2}\right](z) \over c z}\;
	e^{\pi z^2/2\tau_2}\;,
$$
where $c = \vartheta'\left[{1/2 \atop 1/2}\right](0)$.  The
characteristic is the unique odd characteristic in genus~1; in this
case $\kappa = \sqrt{K}$ with $\sqrt{K} = \left[{1/2 \atop
1/2}\right]$, the odd spin structure.  Note that $\detonep(z)$ is a
holomorphic function on $\Hzo$.  As noted earlier, the operator
$\BBox_-(z)$ does not have covariance properties under the lattice and
therefore we do not expect $\detonep(z)$ to be a section of a line
bundle over $J(\Sigma)$.  This is explicitly verified by the formula
above.

\section{Final Results}
\label{final-section}

We can now collate our results and obtain a formula for (\ref{Z-sc}). 
We remind the reader that $d=2n$.  The semiclassical partition
function is given by
$$
	\Zsc = \left(\frac{\vol \Sigma}{2\pi}\right)^{d/2} \int_M 
	\left(\det{}'\dhalf\right)^{d/2}
	\left(\detprime i*(\dbar + \cR \otimes \hat{h}_\delta^2) 
	\partial \right)^{-1/2}
$$
where $h_\delta$ is a holomorphic $(0,1/2)$-spinor for spin structure
$\delta$ and $\hat{h}_\delta$ is a unit normalized spinor.  Note that
we can write $h_{\delta} = N \hat{h}_{\delta}$ where $\langle
h_{\delta},h_{\delta}\rangle = |N|^{2}$.  It well known
\cite{Fay:theta,Mumford:2} that the square of the holomorphic spinor
may be taken to be
\begin{equation}
	h_\delta^2 = \sum_{k=1}^g 
	\overline{\frac{\partial\vartheta[\delta](0)}{\partial z_k}}
	\bar{\omega}_k \;.
	\label{eq:hsq}
\end{equation}
Using our results and assuming $\pont_1(M)=0$ we get
\begin{equation}
	\Zsc = \left(\frac{\vol \Sigma}{2\pi}\right)^{d/2} \int_M 
	\left(\det{}'\dhalf\right)^{d/2} \left(\detp\right)^{-d/2}
	\prod_{\alpha = 1}^n
	{\zanalog(2\pi i x_\alpha \hat{h}_\delta^2) \over
	\vartheta[-\kappa](2\pi i x_\alpha \hat{h}_\delta^2)}\;,
	\label{eq:Zschat}
\end{equation}
where $\kappa$ is the characteristic given by the vector of Riemann
constants.  The Hodge matrix is $H_{jk} = \langle \omega_j, \omega_k
\rangle = (\Omega-\bar{\Omega})_{jk}/2i$.  Also note that
$\hat{h}_\delta^2 = h_\delta^2 / N^{2}$.  The integral over $M$ arises
from the terms which are homogeneous of degree $n$ in the
$x_\alpha$'s.  Using this we  scale  the normalization of the
spinor and obtain
\begin{eqnarray}
	\Zsc &= & 
	\left({\det{}'\dhalf \over N^{2}}\right)^{d/2}
	\left({\vol\Sigma \det H \over \detp}\right)^{d/2} 
	\nonumber \\
	& \times & \left(\frac{1}{2\pi}\right)^{d/2}
	\left(\frac{1}{\det H}\right)^n \int_M \prod_{\alpha = 1}^n 
	{\zanalog(2\pi i x_\alpha h_\delta^2) \over 
	\vartheta[-\kappa](2\pi i x_\alpha h_\delta^2)}\;.
	\label{eq:beauty}
\end{eqnarray}
In terms of the standard 
coordinates $(z_1,\ldots,z_g)$ on $\Hzo$,
$$
	\Azo = 2\pi i \sum_{j,k} z_j (\Omega-\bar{\Omega})_{jk}^{-1} 
	\bar{\omega}_k \;.
$$
The $j$-th coordinate for $2\pi i x_\alpha h_\delta^2$ is
$$
	z_j = x_\alpha \sum_k (\Omega-\bar{\Omega})_{jk}
	\overline{\frac{\partial\vartheta[\delta](0)}{\partial z_k}}\;.
$$
We schematically write $z= x_\alpha(\Omega-\bar{\Omega})
\overline{\vartheta'[\delta](0)}$.  In this notation our formula
becomes
\begin{theorem}
\begin{eqnarray}
	\Zsc &=& 
	\left({\det{}'\dhalf \over N^{2}}\right)^{n}
	\left({\vol\Sigma \det H \over \detp}\right)^{n} 
	\nonumber \\
	&\times& \left(\frac{1}{2\pi}\right)^{n} \left(\frac{1}{\det
	H}\right)^n \int_M \prod_{\alpha = 1}^n
	{\zanalog(x_\alpha(\Omega-\bar{\Omega})
	\overline{\vartheta'[\delta](0)}) \over
	\vartheta[-\kappa](x_\alpha(\Omega-\bar{\Omega})
	\overline{\vartheta'[\delta](0)})}\;,
	\label{second}
\end{eqnarray}
where $\dim M = d = 2n$.  $\Zsc$ is a section of
$\left((\DET(\ddelta))^{*}\right)^{n}\otimes \Sym^{n}\Hzo$ over odd
spin Teichmuller space
\end{theorem}

\begin{description}
	
\item[Remark 1.]
It is important to better understand $\zanalog/\vartheta$ in the
expression for $\Zsc$.  Let $\mathcal{J}$ denote
the holomorphic bundle over $\Mhalf$ whose fiber over a point is the
jacobian\footnote{See Appendix~\ref{app-kappa}.} $J_{0}$ for that
modulus.  Since the point in $\Mhalf$ specifies the spin as well as
the modulus, the square of the holomorphic spinor gives a vector in $\Hoz$ The
determinant line bundle $\DET D$ is a holomorphic line bundle over
$\mathcal{J}$ (although we can forget the spin structure).  The
restriction of $\DET D$ to the zero cross section
$c:\Mhalf\to\mathcal{J}$ obtained by choosing the trivial line bundle
$0\in J_{0}$ is the Hodge line bundle over $\Mod$ pulled up to
$\Mhalf$.

We want to show that the integrand in the second line of
(\ref{second}) is a function in an appropriate ``small'' neighborhood
of the cross section $c(\Mhalf)$.  Choose a point in $\Mhalf$.  We
will see that we can lift a neighborhood $\mathcal{U}$ of this point
to a neighborhood of the covering space of $\Mhalf$ by choosing a
fixed symplectic basis $b$ of $H_{1}(\Sigma,\bbZ)$ which in turn
determines $(\omega_{1},\ldots,\omega_{g})$ in the usual way.  We can
think of the neighborhood $\mathcal{U}$ as a collection of metrics
close to one metric, a choice of symplectic basis $b$ and a choice of
spin structure\footnote{\label{foot-symp} We point out that a choice
of symplectic basis determines a spin structure $S_{b}$ so that our
choice of spin structure $\delta$ is $S_{b}$ plus an element of
$H^{1}(\Sigma,\bbZ_{2})$.  Put another way, a choice of symplectic
basis lifts a small open set in $\Mod$ to spin Teichmuller space
$\mathcal{T}^{1/2}$ as well as to Teichmuller space $\mathcal{T}$. 
Further another choice $b'$ of symplectic basis determined by an
element of $\Sp(g,\bbZ)$ sends $S_{b}$ to $S_{b'}$ and acts on
$H^{1}(\Sigma,\bbZ_{2})$ in the usual way.}.  Choose the neighborhood
$\mathcal{N}_{\mathcal{U}}$ of $c(\mathcal{U})$ to be an open set in
$\mathcal{J}$ for which $(\varpi_{1},\ldots,\varpi_{g})$ makes sense,
i.e., we can solve PDE (\ref{pdevarpi}) that defines the $\varpi_{j}$.

Now $(\varpi_{1}\wedge\cdots\wedge\varpi_{g})^{-1}$ is a holomorphic 
cross section of $\DET D$ restricted to $\mathcal{N}_{\mathcal{U}}$ 
which is $(\omega_{1}\wedge\cdots\wedge\omega_{g})^{-1}$ on 
$c(\mathcal{U})$, a cross section of the Hodge line bundle. We note 
from (\ref{theta-pt}) and (\ref{def-theta-u}) that
\begin{equation}
	\frac{\zanalog(A)}{\vartheta[-\kappa](A)} =
	e^{\pi i \sum z_j (\Omega-\bar{\Omega})^{-1}_{jk} z_k}
	(\varpi_{1}\wedge\cdots\wedge\varpi_{g})
	e^{-\int_0^A(\nutil-\rho)}
	(\omega_{1}\wedge\cdots\wedge\omega_{g})^{-1}
	\label{eq:theta-ratio}
\end{equation}
where we have made the specific choice 
$s_{0}=(\omega_{1}\wedge\cdots\wedge\omega_{g})^{-1}$ so that $r=1$, 
see (\ref{def-r}).  We ignore the holomorphic term $e^{\pi i \sum z_j 
(\Omega-\bar{\Omega})^{-1}_{jk} z_k}$ because when $z$ is substituted 
by $x_{\alpha}h^{2}_{\delta}$ and the product is taken over 
$\alpha=1,\ldots,g$ we get $\pont_{1}(M)$ in the exponent and it 
vanishes under our assumptions.  The term $e^{-\int_0^A(\nutil-\rho)} 
(\omega_{1}\wedge\cdots\wedge\omega_{g})^{-1}$ is parallel transport 
of $(\omega_{1}\wedge\cdots\wedge\omega_{g})^{-1}$ to an element of 
$\DET D$ at $A\in\Hzo$ over the projection of $A$ in $\mathcal{U}$.  
This is a holomorphic cross section of $\DET D$ restricted to 
$\mathcal{N}_{\mathcal{U}}$ because $\dbar(\nutil-\rho)=0$.  The 
quotient of the two terms is a holomorphic function, 
$\mathcal{F}_{\mathcal{N}_{\mathcal{U}}}$ on 
$\mathcal{N}_{\mathcal{U}}$.

We now show that
$\mathcal{F}_{\mathcal{N}_{\mathcal{U}}}=
\mathcal{F}_{\mathcal{N}_{\mathcal{V}}}$ over 
$\mathcal{N}_{\mathcal{U}}\cap\mathcal{N}_{\mathcal{U}}$, if 
$\mathcal{V}$ is another choice of open set in $\mathcal{T}$, which is 
tantamount to another choice of symplectic basis in 
$H_{1}(\Sigma,\bbZ)$. The change of basis is given by an element 
in $\Sp(g,\bbZ)$ represented by the matrix 
$ A \ B \choose C \ D$.
On the overlap $(\omega_{1}\wedge\cdots\wedge\omega_{g})$ is 
transformed into 
$\det(C\Omega+D)(\omega_{1}\wedge\cdots\wedge\omega_{g})$ and 
similarly $(\varpi_{1}\wedge\cdots\wedge\varpi_{g})$ is 
transformed into 
$\det(C\Omega+D)(\varpi_{1}\wedge\cdots\wedge\varpi_{g})$. However 
both $\rho$ and the Quillen connection $\nutil$ are independent of a 
choice of symplectic basis. Hence the two factors in the numerator 
and the denominator in (\ref{eq:theta-ratio}) cancel.

\item[Remark 2.]
Footnote~\ref{foot-symp} explains how $h_{\delta}$ transforms so 
that $h^{2}_{\delta}\in\Hzo$ transforms properly under $\Sp(g,\bbZ)$. 
We need this fact because we evaluate our function on a small multiple 
of $x_{\alpha}h^{2}_{\delta}$ in $\mathcal{N}_{\mathcal{U}}$.

\item[Remark 3.]
We next show that specializing
(\ref{second}) to $g=1$ gives the elliptic genus
\begin{equation}
	\Zsc = \int_{M} \prod_{i=1}^{n} \left[ 
	\frac{ix_{\alpha}}{\vartheta (ix_{\alpha}/2\pi,\tau)} \cdot 
	\eta(q)\right]\,.
	\label{eq:elliptic1}
\end{equation}
Let $q = e^{2\pi i \tau}$ be the modular parameter for the torus.  The
Dedekind eta function is defined by $\eta(q) = q^{1/24}
\prod_{n=1}^\infty (1 - q^n)$.  The theta function
$-\vartheta[-\kappa](z,\tau)$ is the standard odd theta function and
it will simply be denoted by $\vartheta(z,\tau)$.  The spin structure
$\delta$ is the unique odd structure so $\vartheta[\delta](z,\tau) =
\vartheta(z,\tau)$.  We need the identity $\vartheta'(0,\tau) = -2\pi
\eta(q)^{3}$.  Note that the abelian differential is $\omega = dz$,
$h^2_\delta = \overline{\vartheta'(0,\tau)}d\zbar$,
$\zanalog(z)=-\vartheta'(0,\tau) z$, $\langle dz, dz \rangle =
\tau_{2}$, $\langle \sqrt{dz}, \sqrt{dz}\rangle = \tau_{2}$, $\langle
h_\delta, h_\delta \rangle = \tau_{2}|\vartheta'(0,\tau)|$.  We
judiciously choose $N^{2} = -\tau_{2} \vartheta'(0,\tau)$.  We also
have the standard results
\begin{eqnarray*}
	{\det{}'\dhalf \over 2\tau_2} & = & 
	\left(\overline{\eta(q)}\right)^2 \\
	{4 \tau_2^2 \over \detp} & = & \left( \eta(q)\overline{\eta(q)}
	\right)^{-2}\,.
\end{eqnarray*}
Inserting into (\ref{second}) we see that all dependence on
$\bar{\tau}$ disappears and we are left with (\ref{eq:elliptic1}). 
This means that $\Zsc$ is a holomorphic function of $\tau$.

\item[Remark 4.]
We now explain footnote~\ref{foot:Zsc}, \emph{i.e.}, $\Zsc=0$ when
$\dim\ker\dirac>1$.  Let $\ddelta$ denote the $\dirac$ for odd spin
structure $\delta\in\Mhalf$ and let $V_{0}= \{ \delta\in\Mhalf \,|\,
\dim\ker\ddelta =1\}$; $V_{0}$ is an open set in $\Mhalf$.  When
$\delta\in V_{0}$, formula (\ref{second}) holds for $\Zsc(\delta)$. 
We show that $\Zsc(\delta) \to 0$ as $\delta \to \Mhalf - V_{0}$.

First, $\det'\ddelta \to 0$ as $\delta \to \Mhalf - V_{0}$ because
some nonzero eigenvalues of $\ddelta^{*} \ddelta$ must approach $0$ if
$\dim\ker\ddelta$ is to become greater than $1$.  So, looking at
(\ref{eq:Zschat}), we need only to show that $\hat{h}^{2}_{\delta} \in
\Hzo$ remains bounded as $\delta \to \Mhalf - V_{0}$.  But
$\hat{h}^{2}_{\delta} = h^{2}_{\delta}/N^{2}$ and its norm squared in
$\Hzo$ is $\langle h^{2}_{\delta}/N^{2},
h^{2}_{\delta}/N^{2}\rangle_{\Sigma}$.  A Rellich inequality gives
boundedness: the map from the Sobolev space $H^{s}(\Sigma)$ to the
continuous functions $C^{0}(\Sigma)$ is continuous for $s>1$,
\emph{i.e.}, there exists a constant $\kappa>0$ such that $(\sup\norm
f) \le \kappa \Vert f \Vert_{s} = \kappa \langle f, (1 +
\Delta)^{s}f\rangle_{\Sigma}^{1/2} = \kappa \langle f,
f\rangle^{1/2}_{\Sigma}$ if $f$ is harmonic.  This statement is also
true if $f$ is a harmonic form and not just a function.  So we have
$(\sup\norm h^{2}_{\delta}/N^{2})^{2} \le \kappa^{2} \langle
h^{2}_{\delta}/N^{2}, h^{2}_{\delta}/N^{2} \rangle_{\Sigma} \le
\kappa^{2} (\sup\norm h^{2}_{\delta}/N^{2}) \langle h_{\delta}/N,
h_{\delta}/N\rangle_{\Sigma} = \kappa^{2} (\sup\norm
h^{2}_{\delta}/N^{2})$.  Thus we conclude that $(\sup\norm
h^{2}_{\delta}/N^{2}) \le \kappa^{2}$.

\item[Remark 5.]
The factor before the integral sign in $\Zsc(M)$,
equation~(\ref{eq:beauty}), depends only on the Riemann surface and the
dimensionality of the target $M$.  It is difficult to compute.  We can
cancel it by taking ratios for two manifolds of the same dimension. 
Thus
$$
	{\Zsc(M_{1}) \over \Zsc(M_{2})}=
	{\displaystyle 
	\int_{M_{1}} \prod_{\alpha = 1}^n 
	{\zanalog(2\pi i x_\alpha h_\delta^2) \over 
	\vartheta[-\kappa](2\pi i x_\alpha h_\delta^2)}
	\over
	\displaystyle
	\int_{M_{2}} \prod_{\alpha = 1}^n 
	{\zanalog(2\pi i x_\alpha h_\delta^2) \over 
	\vartheta[-\kappa](2\pi i x_\alpha h_\delta^2)}
	}
	=
	{\displaystyle \int_{M_{1}}
	\prod_{\alpha = 1}^n
	{\zanalog(x_\alpha(\Omega-\bar{\Omega})\overline{\vartheta'[\delta](0)})
	\over \vartheta[-\kappa](x_\alpha(\Omega-\bar{\Omega})
	\overline{\vartheta'[\delta](0)})} \over \displaystyle
	\int_{M_{2}} \prod_{\alpha = 1}^n
	{\zanalog(x_\alpha(\Omega-\bar{\Omega})
	\overline{\vartheta'[\delta](0)}) \over
	\vartheta[-\kappa](x_\alpha(\Omega-\bar{\Omega})
	\overline{\vartheta'[\delta](0)})}} \;.
$$

The formula above can be made more explicit by a judicious choice of
$M_{2}$.  Kervaire and Milnor \cite{Kervaire,Hirzebruch:modular}
constructed a manifold $M_{0}^{4k}$ of dimension $4k$ whose only
nonvanishing class is $\pont_{k}$.  If $\dim M_{1} = 8k$ then let
$M_{2} = (M_{0}^{8})^{k}$; if $\dim M_{1}= 8k+4$ then let $M_{2}=
(M_{0}^{8})^{(k-1)}\times M_{0}^{12}$.  With these choices of $M_{2}$
the integral in $\Zsc$ is easy to compute (for manifolds of low
dimension) using the first few terms in the power series expansion of
$\zanalog/\vartheta$.

\item[Remark 6.]
We have constructed a section $\Zsc$ over over odd spin Teichmuller
space but we do not know if it gives a section over odd spin moduli
space $\Mhalf$. 

\end{description}

% Put appendices below \appendix command
\appendix

\section{Identifying a holomorphic cross section of $\cL$
with a $\vartheta$ function}
\label{app-kappa}

At the beginning of Section~\ref{sec-diff} we stated that the 
operator $D = * i (\dbar + \Azo)$ varied holomorphically on $\Hzo$ 
and because of equivariance defined a family of operators 
parametrized by $J(\Sigma)$. This family is in fact $\dbar$ twisted 
by flat line bundles as we shall see. Later we needed its determinant 
line bundle and we needed to identify a holomorphic cross section 
(unique up to scale) with the appropriate $\vartheta$-function.

In this appendix we make the identification using facts about Riemann 
surfaces well known to algebraic geometers.

Let $J_r(\Sigma)$ denote the set of holomorphic line bundles over 
$\Sigma$ with $\cclass_1$ equal to $r$. Of particular interest to us 
is $J_0(\Sigma)$ (our $J(\Sigma)$), the set of flat line bundles, 
which we can identify with $\widetilde{\pi_1(\Sigma)} = \{ 
\chi:\pi_1(\Sigma) \to S^1\} \simeq \widetilde{H_1(\Sigma,\bbZ)}
\simeq H^1(\Sigma,\bbR)/H^1(\Sigma,\bbZ)$. The last isomorphism can 
be described in terms of (real) closed $1$-forms $\omega$: let 
$\chi_\omega(\gamma) = \exp\left(2\pi i \int_\gamma \omega\right)$ 
for $\gamma$ a closed path starting at $P_0\in\Sigma$. Clearly 
$\chi_\omega$ is a homomorphism $\pi_1(\Sigma)\to S^1$ and depends 
only on the cohomology class of $\omega$. It is easy to see that 
$\omega \mapsto \chi_\omega$ induces an isomorphism 
$H^1(\Sigma,\bbR)/H^1(\Sigma,\bbZ) \to \widetilde{\pi_1(\Sigma)}$. 
Here we take $\pi_1(\Sigma)$ as the closed (piecewise smooth) paths 
starting at $P_0$ with equivalence relationship given by homotopy.

When the real surface $\Sigma$ has a complex structure, then
$H^1(\Sigma,\bbC) = H^1(\Sigma,\bbR)\otimes\bbC \simeq \Hoz \oplus
\Hzo$.  Taking the real part induces an isomorphism of $\Hzo$ (or
$\Hoz$) with $H^1(\Sigma,\bbR)$.  Let $\real: \Hzo \to
H^1(\Sigma,\bbR)$ be this isomorphism, and let $\widetilde{\Hzo} =
(\real)^{-1} H^1(\Sigma,\bbZ)$ so that
$H^1(\Sigma,\bbR)/H^1(\Sigma,\bbZ) \simeq \Hzo/\widetilde{\Hzo}$. 
Since $\Hzo$ is a complex vector space, $\Hzo/\widetilde{\Hzo}$ is a
complex torus $J(\Sigma)$, the jacobian of $\Sigma$.  Our chain of
arguments demonstrates that the jacobian $J_0(\Sigma)$ is isomorphic
to $\Honehat$, the character group of $H_{1}(\Sigma,\bbZ)$. 
Specifically, if $\mu \in \Hzo$, let
\begin{equation}
	\chi_\mu(\gamma) = e^{2\pi i \int_\gamma(\mu 
	+\bar{\mu})/2}\;,
	\label{chi-mu}
\end{equation}
with $\gamma$ a loop with basepoint $P_0$, then 
$\mu\to \chi_\mu$ induces the isomorphism of $\Hzo/L_\Omega = 
J(\Sigma)$ with $\Honehat$. The covering space of 
$J(\Sigma)=J_0(\Sigma)$ is $\Hzo$.

In Sections~\ref{sec-diff} and \ref{hol-triv} we chose a standard 
basis $(\bar{\omega}_{1},\ldots,\bar{\omega}_{g})$ of $\Hzo$ obtained 
from a choice of symplectic basis $\{A_{i},B_{j}\}$ of 
$H_{1}(\Sigma,\bbR)$ so that $\int_{A_{i}}\omega_{j}=\delta_{ij}$. We 
remind the reader that the Riemann period matrix $\Omega_{ij}= 
\int_{B_{i}}\omega_{j}$ with imaginary part 
$(\Omega_{ij}-\bar{\Omega}_{ij})/2i$ that is positive definite and in fact 
equal to $\langle \omega_{i},\omega_{j}\rangle$.
Then $\omega_i = \sum\alpha_i + 
\Omega_{ij} \beta_j$ where $\alpha$ and $\beta$ are the harmonic 
representatives dual to the $a$ and $b$ cycles.  We write a point in 
$\Hzo$ as $(u_j + i v_j)\bar{\omega}_j$ where $u$ and $v$ are 
real.  One can easily verify that $\widetilde{\Hzo}$ is represented by 
$2\sum(m + \Omega n)_j(\Omega-\bar{\Omega})_{jk}^{-1}\bar{\omega}_k$ 
where $m \in\bbZ^g$ and $n\in\bbZ^g$.  This is not quite 
(\ref{B-coord}).  The discrepancy arises because of the factors of 
$\pi$ and $i$ in the exponent of (\ref{chi-mu}). Taking these factors 
into account leads to the normalization (\ref{z-coord}) and 
(\ref{B-coord}). When we write $\vartheta(\Azo)$ we mean 
$\vartheta(z)$ where $z$ is given by (\ref{z-coord}). The 
quasiperiodicity properties of the theta function are associated with 
$z \to z + (m + \Omega n)$.

Multiplication $m_L$ by a line bundle $L \in J_r(\Sigma)$ gives an 
isomorphism $m_L : J_0(\Sigma) \to J_r(\Sigma)$. In particular a spin 
structure $\sqrt{K}$, where $K$ is the canonical bundle of $\Sigma$, 
gives $m_{\sqrt{K}}: J_0(\Sigma) \to J_{g-1}(\Sigma)$ with $g$ the 
genus of $\Sigma$. Similarly for $P_0\in\Sigma$ let $\Lpz$ be the 
line bundle with divisor $P_0$ so that $\Lpz \in J_1(\Sigma)$. Then 
$m_{\Lpz^r}: J_0(\Sigma) \to J_r(\Sigma)$ is an isomorphism. The 
complex structure on $J_r$ is chosen such that $m_L$ is holomorphic.

One can construct a Poincar\'{e} line bundle $Q_r$ over 
$J_r(\Sigma)\times\Sigma$ whose restriction to each fiber $\{L\}\times 
\Sigma$ is the line $L\in J_r(\Sigma)$.  The holomorphic line bundle 
$Q_r$ over $J_r(\Sigma)\times\Sigma$ is determined only up to a line 
bundle on $J_r(\Sigma)$ pulled up to $J_r(\Sigma)\times\Sigma$. A 
choice of point $P_0\in\Sigma$ determines $Q_r$ by stipulating that 
$Q_r|_{J_r(\Sigma)\times\{P_0\}} \simeq 1$ on $J_r(\Sigma)$.
In Appendix~\ref{app-Q} below we construct such a $Q_0$ explicitly. 
We can use $(m_{L_{P_{0}}^{g-1}})^{*} Q_{g-1}$ instead.

Let $\dbar\otimes I_{Q_r}$ be the family of $\dbar$ operators 
parametrized by $Q_r$.  Suppose $\cM$ is a holomorphic line bundle on 
$J_r(\Sigma)$ which pulled up to $J_r(\Sigma)\times\Sigma$ is $\cMt$.  
Suppose we have modified our choice of Poincar\'{e} line bundle $Q_r$ 
by $Q_r\otimes\cMt$.  One can show the determinant line bundle of the 
family $\dbar\otimes I_{Q_r\otimes\cMt}$, $\DET(\dbar\otimes 
I_{Q_r\otimes\cMt})$ is isomorphic to $\DET(\dbar\otimes I_{Q_r})
\otimes \cM^{r+1-g}$.  In particular, when $r=g-1$, $\DET(\dbar\otimes 
I_{Q_{g-1}})$ is independent of choice of $Q_{g-1}$.

The choice of $r=g-1$ is special because the index of the operator
$\dbar\otimes I_L$, $L\in Q_{g-1}$, is zero.  Generically the operator
$\dbar\otimes I_L$ is invertible.  Let $\cV = \{ L\in J_{g-1}(\Sigma)
\;|\; \dbar\otimes I_L \quad\mbox{is not invertible}\}$.  $\cV$ is a
variety in $J_{g-1}(\Sigma)$, in fact the divisor of the line bundle
$\DET(\dbar\otimes I_{Q_{g-1}})$.  Of course, $\cV$ is also $\{ L\in
J_{g-1}(\Sigma) \;|\; L \quad\mbox{has a nonzero holomorphic
section}\}$.

We use $m_{L^{g-1}_{P_0}}$ to compare $\DET(\dbar\otimes I_{Q_{g-1}})$ 
with $\DET(\dbar\otimes I_{Q_0})$, the latter our line bundle $\cL$ 
over $J_0(\Sigma)$. The Grothendieck-Riemann-Roch theorem implies that
$$
	\left(m_{L^{g-1}_{P_0}}\right)^*\left(\DET(\dbar \otimes 
	I_{Q_{g-1}})\right)
$$
is isomorphic to $\cL$.  Although it is well known that 
$H^0(\cL,\bbC)$ has complex dimension one, i.e., the holomorphic 
sections of $\cL$ form a one dimensional subspace, we explain this in 
Appendix~\ref{app-holosec}.  We now want to identify a properly 
normalized holomorphic section of $\cL$ with a $\vartheta$-function.

First  identify $\widetilde{J_0(\Sigma)}$, 
the universal cover of $J_0(\Sigma)$, with $\bbC^g$, where we have 
defined 
the Riemann theta function $\vartheta$ and its divisor. Let 
$\cL_\vartheta$ be the holomorphic line bundle over $J_0(\Sigma)$ 
whose divisor pulls up to the divisor of $\vartheta$. We learn from 
Riemann surface theory that there exists a spin structure 
$\sqrt{K}\in J_{g-1}(\Sigma)$ such that $\cL_\vartheta = 
m^*_{\sqrt{K}} \DET(\dbar\otimes I_{Q_{g-1}})$. Put another way, 
$m_{\sqrt{K}}(\mbox{divisor of $\cL_\vartheta$}) = \cV$. The spin 
structure is the one determined by the choice of symplectic basis of 
cycles in $H_{1}(\Sigma,\bbR)$.

Putting these two facts together gives $\cL \simeq
\left(m^{-1}_{\sqrt{K}} m_{L^{g-1}_{P_0}}\right)^* \cL_\vartheta
\simeq \left( m_{K^{-1/2} L^{g-1}_{P_0}}\right) \cL_\vartheta$.  The
flat line bundle $K^{-1/2} L^{g-1}_{P_0}$ lies in $J_0(\Sigma)$ and is
in fact $-\kappa$ where $\kappa$ is the Riemann constant 
\cite[p. 338]{Griffiths:Harris}.  Hence $\cL$ is the translate of
$\cL_\vartheta$ by $-\kappa$.  As a result, our $\vartheta\left[{a
\atop b}\right]$ is the translate of $\vartheta$ by $-\kappa$; the
characteristic $\left[{a \atop b}\right]$ equals $-\kappa$.

\section{Explicit construction of $Q_0$}
\label{app-Q}

We now construct $Q_0$ and identify the family $\dbar\otimes I_{Q_0}$ 
with the equivariant family $*i(\dbar + \Azo)$, $\Azo\in\Hzo$.

Fix a point $P_0\in\Sigma$ once and for all, and construct the simply 
connected covering space $\Sigmat$ of $\Sigma$ as the (piecewise 
smooth) space of equivalence classes of paths starting at $P_0$ with 
equivalence the relation given by homotopy.  $\Sigmat$ is a principal 
bundle over $\Sigma$ with group $\pi_1(\Sigma) = \pi_1(\Sigma,P_0)$ 
(the closed paths starting at $P_0$) and projection map the endpoint 
map.

The Poincar\'{e} line bundle $Q$ is the complex line bundle over
$J(\Sigma)\times \Sigma$ defined as follows: for each $\chi\in
J(\Sigma)$, $Q_\chi$ over $\Sigma$ is $\Sigmat\times_\chi \bbC$ with
$\pi_1(\Sigma)$ acting on $\bbC$ via $\chi$.  Standard arguments
demonstrate (see for example \cite{Griffiths:Harris}) that
$\{Q_\chi\}_{\chi\in J(\Sigma)}$ can be made into a holomorphic line
bundle over $J(\Sigma)\times \Sigma$ with hermitian metric (because
$\chi:\pi_1(\Sigma) \to S^1$).  One can think of this as a family of
line bundles parametrized by $J(\Sigma)$.  Note that
$Q|_{J(\Sigma)\times\{P_0\}}$ is $J(\Sigma)\times\bbC$ because for any
$\chi$, $Q_\chi$ at $(\chi,P_0)$ is $\{P_0\}\times\bbC$ with $\{P_0\}$
the constant path at $P_0$.  For each $\chi$, $Q_\chi$ is a
holomorphic flat line bundle over $\Sigma$ with hermitian metric hence
comes equipped with a unique $(1,0)$ connection $a_\chi$ on the
associated $\bbC^*$ bundle over $\Sigma$; moreover, $\dbar a_\chi=0$
since $Q_\chi$ is flat\footnote{Of course, this connection is also the
one which is $0$ on the local section of $\Sigma \to \Sigmat$.}.

We now trivialize $Q_\chi$ by finding a $C^\infty$ nonvanishing 
section. Since $Q_\chi = \Sigmat \times_\chi \bbC$, a section of 
$Q_\chi$ is a complex valued function $f$ on $\Sigmat$ such that 
$f(\tilde{\sigma}\cdot \gamma) = \chi(\gamma) f(\tilde{\sigma})$ for 
$\gamma$ a loop based at $P_0$. Suppose $\chi = \chi_\omega$ for 
$\omega\in\Hzo$, i.e., $\chi_\omega(\gamma) = e^{\pi i 
\int_\gamma(\omega + \bar{\omega})}$. Define $f(\tilde{\sigma}) = 
e^{\pi i \int_{\tilde{\sigma}} (\omega+\bar{\omega})}$, remembering 
that $\tilde{\sigma}$ is a path with $\tilde{\sigma}(0)=P_0$.  Since
$\int_{\tilde{\sigma}\cdot\gamma}(\omega+\bar{\omega}) =
\int_{\tilde{\sigma}}(\omega+\bar{\omega}) +
\int_{\gamma}(\omega+\bar{\omega})$, $f$ is a nonvanishing section 
of $Q_\chi$.

\begin{figure}[tbp]
	\centering
	\includegraphics{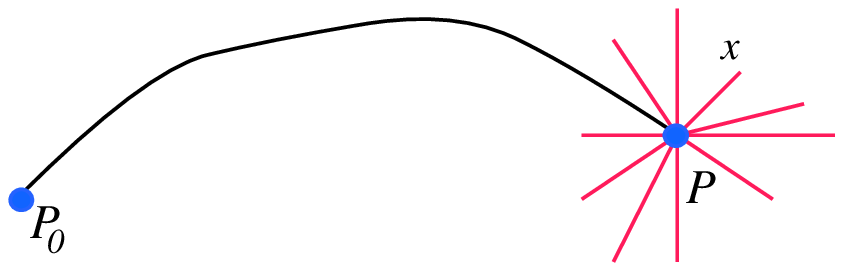}
	\caption[xyz1]{The lift of a coordinate neighborhood $N$ of $P$
	into $\Sigmat$.}
	\label{fig:lift}
\end{figure}

$f$ also gives a section $\tilde{f}$ of the circle bundle $P_\chi = 
\Sigmat \times_\chi S^1$ over $\Sigma$ because $f$ has values in 
$S^1$.  Hence $\tilde{f}$ identifies $P_\chi$ with $\Sigma\times S^1$ 
by $(\gamma, e^{2\pi i\theta}) \mapsto (\gamma(1), f(\gamma) e^{2\pi 
i\theta})$; thus the canonical connection on $P_\chi$ becomes a 
$1$-form on $\Sigma = \Sigma \times \{1\} \hookrightarrow \Sigma 
\times S^1$ which we now compute.  Given a point $P\in \Sigma$ we have 
the lift of a coordinate neighborhood $N$ of $P$ into $\Sigmat$.  
Figure~\ref{fig:lift} explains the lift.  
We choose a path $\gamma_P$ from $P_0$ to $P$ and follow it by the 
straight line from $P$ to $x$.  Call $\gamma_x$ this path from $P_0$ 
to $x$.  Then $x \mapsto \gamma_x$ is a lift of the coordinate 
neighborhood to a neighborhood about $\gamma_P$.  The connection 
$1$-form on $P_\chi$ is $0$ on $(\gamma_x,1)$ and so is $d\theta$ on 
$(\gamma_x,e^{2\pi i \theta})$.  Now $\tilde{f}(\gamma_x,e^{2\pi i 
\theta}) = (x, e^{\pi i \int_{\gamma_x}(\omega+\bar{\omega})} e^{2\pi 
i \theta})$.  Hence the $1$-form on $\Sigma$ is 
$\half(\omega+\bar{\omega})$ in the trivialization we used.  With our 
choices the $(0,1)$ part of the connection is 
$\half(I+iJ)(\omega+\bar{\omega})= \omega$, which we denote by $\Azo$ 
in Section~\ref{sec-diff}.

We thus have a family of $\dbar$ operators parametrized by 
$J(\Sigma)$, namely $\{\dbar\otimes I_{Q_\chi}\}_{\chi\in J(\Sigma)}$ 
with canonical connection on $Q_\chi$. When we trivialize $Q_\chi$ as 
above, $\dbar\otimes I_{Q_\chi}$ becomes $\dbar + \omega = \dbar + 
\Azo$. Another trivialization would transform $\omega$ by
$L_\Omega + \dbar \Lambda^0(\Sigma)$ so that if we choose a metric on 
$\Sigma$ and choose $\Hzo$ as the harmonic forms of type $(0,1)$, 
then $\omega$ is determined up to translation by $L_\Omega$ as expected. 
The identification $\{\dbar\otimes I_{Q_\chi}\}_{\chi\in J(\Sigma)}$ 
with $\{ \dbar + \Azo \}_{\Azo\in\Hzo}$ works for each $\chi$ on at 
most a fundamental domain in $L_\Omega$.

With any trivialization $Q_\chi \simeq \Sigma \times \bbC$, the
connection $1$-form $a_\chi$ becomes a $1$-form $A$ on $\Sigma$ which is 
closed because the curvature is $0$. Furthermore, $\chi(\gamma) = 
\exp\left( 2\pi i \int_\gamma A\right)$ so that $A$ is determined as 
a closed $1$-form up to integral $1$-cocycles. Since we have not 
specified the trivialization, $A$ is determined only up to a gauge 
transformation, i.e., as an element of $H^1(\Sigma,\bbR)/H^1(\Sigma, 
\bbZ)$. We could choose $A =(\Azo + \Aoz)/2$ with $\Azo$ the 
harmonic representative in $\Hzo$. Choosing another representative in 
the coset $\Azo + L_\Omega$ amount to choosing another trivialization 
of $Q_\chi$. In any case, $\dbar\otimes I_{Q_\chi}$ becomes $\dbar 
+\Azo$.

We conclude this appendix with the following theorem about the operator $D_A = 
*i(\db + \Azo) : \Loz \to \Lzz$.
\begin{theorem}
	Let $\chi \in J(\Sigma)$ be a character and let $\Azo$ be the 
	associated flat connection.  If $\chi = 1$ then $\dim \ker D_0 = 
	g$.  If $\chi \neq 1$ then $\dim \ker D_A = g-1$.
\end{theorem}
Proof: Since $\ker D_0 = \Hoz$ we have $\dim\ker D_0 = g$. Now the 
index of $D_A$ equals $\Index D_0 = g-1$. To complete the proof of the 
theorem we need only show that $\dim\ker D_A^* = \dim\coker D_A = 0$. 
But $D_A^* = \dbar + \Azo: \Lzz\to\Lzo$ and we have shown that 
$\dbar+\Azo$ is $\dbar: L_\chi \to \Lzo\otimes L_\chi$, for some 
character $\chi$. Now $\cclass_1(L_\chi)=0$ because $L_\chi$ is flat. 
So any holomorphic section of $L_\chi$ has no zeroes. Thus a 
nonvanishing holomorphic section of $L_\chi$ would give a holomorphic 
isomorphism with $\Sigma\times\bbC$ which is not possible if $\chi\neq 
1$. Hence $\dim\ker D^*_A = 0$.

\section{Holomorphic sections of the determinant line bundle}
\label{app-holosec}

We now compute $\cclass_1$ of $\DET(\dbar\otimes I_{Q_0})$ and show 
that the space of holomorphic sections is one dimensional.

We want to compute the determinant line bundle 
$\DET(\dbar\otimes I_Q) \to J(\Sigma)$ of the family
$\{\dbar\otimes I_{Q_\chi}\}_{\chi\in J(\Sigma)}$. We first compute 
its first Chern class using Riemann-Roch or the families index theorem.
\begin{theorem}
	$\cclass_1\left(\DET(\dbar\otimes I_Q)\right)$ is represented by 
	the basic K\"ahler form on $J(\Sigma)$.
\end{theorem}
{\em Proof:} The Chern character of the index bundle over $J(\Sigma)$ 
is
\begin{equation}
	\int_\Sigma \ch(Q) \Todd T(\Sigma)
	\label{ch-L}
\end{equation}
where $T(\Sigma)$ is the vector bundle over $\Sigma\times J(\Sigma)$ 
which is $\Toz$ and is independent of $J(\Sigma)$.
Its Todd class is $1 + \cclass_1(\Toz)/2$. Following convention we 
define $\cclass_1(\Sigma) = \cclass_1(\Toz)$. Further $\ch(Q) = 
\sum_{n=1}^\infty \cclass_1(Q)^n/n!$.

Since the first Chern class of the family is given by the $2$-form 
term in (\ref{ch-L}), we want the $4$-form in the integrand which is 
of total degree~$2$ in the $\Sigma$ direction. If we write 
$\cclass_1(Q)$ as $\mu_{2,0} + \mu_{1,1} + \mu_{0,2}$, the 
decomposition of $\cclass_1(Q)$ along $\Sigma$, $\Sigma\times 
J(\Sigma)$ and $J(\Sigma)$ respectively, we find the desired $4$-form 
to be $\half \mu_{1,1}^2 + \mu_{0,2}\mu_{2,0} + \half\mu_{0,2} 
\cclass_1(\Sigma)$. Since $Q_\chi$ is flat, 
$\mu_{2,0}=\cclass_1(Q_\chi) =0$; so 
$$
	\cclass_1\left(\DET(\dbar\otimes I_Q)\right) = \half \int_\Sigma \left(
		\mu_{1,1}^2 + \mu_{0,2} \cclass_1(\Sigma) \right)\;.
$$

We now compute $\mu_{1,1}$ and show that $\mu_{0,2}=0$. We do so using 
the fact that $\cclass_1(Q) = \frac{1}{2\pi i} \partial\dbar \| F\|^2$ 
where $F$ is a local holomorphic section of $Q$ and $\| \bullet \|$ is 
the norm for the hermitian metric on $Q$.

Let $N_P$ be a neighborhood of $P \in \Sigma$, and let $U_\chi$ be a 
coordinate neighborhood of $\chi\in J(\Sigma)$. Because $\Hzo$ 
covers $J(\Sigma)$, we can identify $U_\chi$ with a neighborhood of 
the appropriate $\Azo \in \Hzo$. In fact we can take $U_\chi = \Azo + 
U_0$ where $U_0$ is a neighborhood of the origin in $\Hzo$.

Fix a path $\gamma_0$ from $P_0$ to $P$. If $x \in N_P$, let 
$\gamma_x$ be the path from $P_0$ to $x$ which is $\gamma_0$ followed 
by the straight line from $P$ to $x$ (see Figure~\ref{fig:lift}).
For $(x,\omega) \in N_P \times U_\chi$ let $F(x,\omega) = \exp^{2\pi i 
\int_{\gamma_x} \omega}$.  For each $\omega \in U_\chi$, $F$ is 
actually a function on the lift of $N_P$ to $\Sigmat$; it 
transforms correctly under $\pi_1(\Sigma)$ to give a local section of 
$Q_{\chi_\omega}$.  We leave the reader to check that $F$ is a 
holomorphic section of $Q$ over $N_P \times U_\chi$.

Now $\frac{1}{2\pi i} \log 
\|F\|^2 (x,\omega) = \int_{\gamma_x} (\omega - \bar{\omega})$ and 
$d_\Sigma$ of this expression is $(\omega - \bar{\omega})$.  Hence 
$\frac{1}{2\pi i}\dbar_{\Sigma} \log \|F\|^2= \omega$ and 
$\frac{1}{2\pi i}\partial_{\Sigma} \log \|F\|^2= -\bar{\omega}$.  
Since $\log \|F\|^2$ is (real) linear in $\omega$,
$$
	\frac{1}{2\pi i}\left(d_\Sigma \log \|F\|^2\right)(\tau) = \int_{\gamma_x} 
	(\tau-\bar{\tau}) \;,
$$
for $\tau \in T_\omega(J(\Sigma)) \cong \Hzo$. Hence
$$
	\frac{1}{2\pi i}\left(\dbar_{\Sigma} \log \|F\|^2\right)(\tau) = 
	-\int_{\gamma_x} \bar{\tau}\;,
$$
while
$$
	\frac{1}{2\pi i}\left(\partial_\Sigma \log \|F\|^2\right)(\tau) = 
	\int_{\gamma_x} \tau\;.
$$
Furthermore linearity implies $\partial_{J(\Sigma)}\dbar_{J(\Sigma)} 
\log \|F\|^2 =0$.  We conclude that $\mu_{0,2}=0$ and 
$\mu_{1,1}(\partial/\partial\zbar,\tau) = 
\tau(\partial/\partial\zbar)$ and $\mu_{1,1}(\partial/\partial z,\tau) 
= -\bar{\tau}(\partial/\partial z)$.  Hence for $\tau \in 
T_\omega(J(\Sigma))$. $\mu_{1,1}(\bullet,\tau)$ is the $1$-form on 
$\Sigma$ equal to $(\tau-\bar{\tau})$ and the two form $\half 
\int_\Sigma \mu_{1,1}^2$ on $J(\Sigma)$ evaluated at $\tau_1$, 
$\tau_2$ equals
$$
	\int_\Sigma (\tau_1 - \bar{\tau}_1) \wedge (\tau_2 - \bar{\tau}_2)
	= \half \int_\Sigma \left(-\tau_1 \wedge \bar{\tau}_2 + \tau_2 \wedge 
	\bar{\tau}_1 \right)\;.
$$
Finally we recall that the metric on $J(\Sigma)$ is given by the inner 
product on $\Hzo$:
$$
	\left\langle \tau_2, \tau_1 \right\rangle =
	\int_\Sigma \tau_1 \wedge *\bar{\tau}_2 = i \int_\Sigma \tau_1 
	\wedge \bar{\tau}_2 \;.
$$
Thus $\half \int_\Sigma \mu_{1,1}^2 = \frac{i}{2} 
(\langle\tau_2,\tau_1\rangle - 
\overline{\langle\tau_2,\tau_1\rangle})$.  Moreover, 
$(\langle\tau_2,\tau_1\rangle - 
\overline{\langle\tau_2,\tau_1\rangle}) = 
-2i\omega(\tau_1,\tau_2)$ with $\omega$ the basic K\"ahler 
form on $J(\Sigma)$.  Hence the first Chern class $\cclass_1\left(\DET(
\dbar\otimes I_Q)\right)$ of the determinant line bundle for $\{\dbar \otimes 
I_{Q_\chi}\}_{\chi\in J(\Sigma)}$ equals $\omega$.

\begin{corollary}
	The space of holomorphic sections of the determinant line bundle
	$\DET(\dbar\otimes I_Q) \to J(\Sigma)$ has dimension~1.
\end{corollary}
{\em Proof:} Because $\cclass_1(\DET(\dbar\otimes I_Q))=\omega$,
$\DET(\dbar\otimes I_Q) \to J(\Sigma)$ is a positive line bundle. The 
Kodaira vanishing theorem applies and tell us that
$\dim H^0(\dbar\otimes I_Q)$ equals the Euler class of the elliptic 
complex
$$
	\Lambda^{0,q} \otimes \DET(\dbar\otimes I_Q)
	\stackrel{\dbar\otimes I_Q}{\longrightarrow}
	\Lambda^{0,q+1} \otimes \DET(\dbar\otimes I_Q)
$$
and is computable by Riemann-Roch. Note that $\Todd J(\Sigma)=1$ hence
$$
	\dim H^0(\dbar\otimes I_Q) = \int_{J(\Sigma)} \ch(Q) \Todd J(\Sigma)
	= \int_{J(\Sigma)} \omega^g/g! = 1\;.
$$

% \bibliographystyle{utphys}
% \bibliography{oabib}
\providecommand{\href}[2]{#2}\begingroup\raggedright\endgroup

\end{document}